\documentclass[aps,prd,notitlepage,showpacs,nofootinbib,superscriptaddress]{revtex4-1}

\usepackage{amsfonts}
\usepackage{amsmath}
\usepackage{amssymb}
\usepackage[english]{babel}
\usepackage{graphicx}
\usepackage{epsfig}
\usepackage{bm}
\usepackage{verbatim}
\usepackage[utf8]{inputenc}
\usepackage{booktabs}
\usepackage{multirow}
\usepackage{caption,subcaption}
\usepackage{xcolor}
\usepackage[colorlinks=true,urlcolor=red,citecolor=red]{hyperref}
\usepackage[font=small]{caption}
\usepackage{float}
\usepackage{blindtext}

\newcommand{\va}{\varphi}

\newcommand{\bc}{\begin{center}}
\newcommand{\ec}{\end{center}}

\newcommand{\si}{\sigma}

\newcommand{\mathsym}[1]{}

\newcommand{\be}{\begin{equation}}
\newcommand{\ee}{\end{equation}}
\newcommand{\bea}{\begin{eqnarray}}
\newcommand{\eea}{\end{eqnarray}}
\input{tcilatex}
\begin{document}

\title{Fermion spectrum and $g-2$ anomalies in a low scale 3-3-1 model}
\author{A. E. C\'arcamo Hern\'andez$^a$}
\email{antonio.carcamo@usm.cl}
\author{Yocelyne Hidalgo Vel\'{a}squez$^a$}
\email{yocehidalgov@gmail.com}
\author{Sergey Kovalenko$^b$}
\email{sergey.kovalenko@usm.cl}
\author{H. N. Long$^{c,d}$}
\email{hoangngoclong@tdtu.edu.vn}
\author{Nicol\'{a}s A. P\'{e}rez-Julve$^a$}
\email{nicolasperezjulve@gmail.com}
\author{V. V. Vien$^{e,f}$}
\email{vvvien@ttn.edu.vn}
\date{\today }

\affiliation{$^a$Universidad T\'{e}cnica Federico Santa Mar\'{\i}a and Centro Cient\'{\i}fico-Tecnol\'{o}gico de Valpara\'{\i}so, \\
Casilla 110-V, Valpara\'{\i}so, Chile\\
$^{b}$ Departamento de Ciencias F\'isicas, Universidad Andres Bello, \\
Sazi\'e 2212, Piso 7, Santiago, Chile\\
$^c$ Theoretical Particle Physics and Cosmology Research Group,\\
Advanced Institute of Materials Science, Ton Duc Thang University, Ho Chi
Minh City, Vietnam\\
$^d$Faculty of Applied Sciences, Ton Duc Thang University, Ho Chi Minh City,
Vietnam\\
$^e$ Institute of Research and Development, Duy Tan University, 182 Nguyen
Van Linh, Da Nang City, Vietnam\\
$^f$ Department of Physics, Tay Nguyen University, 567 Le Duan, Buon Ma
Thuot City, DakLak, Vietnam}

\begin{abstract}
We propose a renormalizable theory based on the $SU(3)_C\times SU(3)_L\times
U(1)_X$ gauge symmetry, supplemented by the spontaneously broken $U(1)_{L_g}$
global lepton number symmetry and the $S_3 \times Z_2 $ discrete group,
which successfully describes the observed SM fermion mass and mixing
hierarchy. In our model the top and exotic quarks get tree level masses,
whereas the bottom, charm and strange quarks as well as the tau and muon
leptons obtain their masses from a tree level Universal seesaw mechanism
thanks to their mixing with charged exotic vector like fermions. The masses
for the first generation SM charged fermions are generated from a radiative
seesaw mechanism at one loop level. The light active neutrino masses are
produced from a loop level radiative seesaw mechanism. Our model
successfully accommodates the experimental values for electron and muon
anomalous magnetic dipole moments.
\end{abstract}

\maketitle

%\title{Fermion masses and mixings and $g-2$ anomalies in a low scale 3-3-1
%model.}

%\email{wvienk16@gmail.com}

\section{\label{intro}Introduction}

Despite of the excellent agreement of the Standard Model (SM) predictions
with the experimental data, there are several %unaddressed issues
problems that do not find explanation within its framework. Among them are
the observed pattern of SM fermion masses and mixing angles, the tiny values
of the light active neutrino masses, the number of SM fermion families, the
electric charge quantization and the anomalous magnetic moments of the muon
and electron. Addressing these issues requires to consider extensions of the
SM with enlarged particle content and symmetries. In particular, theories
based on the $SU(3)_C\times SU(3)_L\times U(1)_X$ gauge symmetry (3-3-1
models) \cite%
{Georgi:1978bv,Valle:1983dk,Pisano:1991ee,Foot:1992rh,Frampton:1992wt,Hoang:1996gi, Hoang:1995vq,Foot:1994ym,CarcamoHernandez:2005ka,Dong:2010zu,Dong:2010gk,Dong:2011vb,Benavides:2010zw,Dong:2012bf,Huong:2012pg,Giang:2012vs,Binh:2013axa,Hernandez:2013mcf,Hernandez:2013hea,Hernandez:2014vta,Hernandez:2014lpa,Kelso:2014qka,Vien:2014gza,Phong:2014ofa,Phong:2014yca,Boucenna:2014ela,DeConto:2015eia,Boucenna:2015zwa,Boucenna:2015pav,Benavides:2015afa,Hernandez:2015tna,Hernandez:2015cra,Hue:2015fbb,Hernandez:2015ywg,Fonseca:2016tbn,Vien:2016tmh,Hernandez:2016eod,Fonseca:2016xsy,Deppisch:2016jzl,Reig:2016ewy,CarcamoHernandez:2017cwi,CarcamoHernandez:2017kra,Hati:2017aez,Barreto:2017xix,CarcamoHernandez:2018iel,Vien:2018otl,Dias:2018ddy,Ferreira:2019qpf,CarcamoHernandez:2019vih,CarcamoHernandez:2019iwh,CarcamoHernandez:2019lhv,CarcamoHernandez:2020pnh}%
, have received a lot of attention since they answer some of the open
questions of the SM, such as, for example, the number of SM fermion families
and the electric charge quantization. Adding discrete symmetries and
extending the scalar and fermionic content of such 3-3-1 models allows
addressing the observed SM fermion mass and mixing hierarchy. Furthermore,
if one considers 3-3-1 models where the fermions do not have exotic electric
charges, the third component of the $SU(3)_L$ leptonic triplet will be
electrically neutral. This allows the implementation of a low scale linear
or inverse seesaw mechanism producing the tiny light active neutrino masses
and sterile neutrinos with masses at the $SU(3)_L\times U(1)_X$ symmetry
breaking scale, thus making the model testable at colliders.

Imposing discrete symmetries allows one to forbid tree level masses arising
from the Standard Yukawa interactions for the SM fermions lighter than the
top quark. To generate such masses, one has to consider heavy vector-like
fermions, mixing with the SM fermions lighter than the top quark, as well as
gauge singlet scalar fields. Their inclusion in the particle spectrum of the
model is crucial for the implementation of the Universal and radiative
seesaw mechanisms needed to generate the masses for the SM fermions lighter
than the top quark, thus explaining the SM charged fermion mass hierarchy.
In addition, the heavy vector like leptons can provide an explanation for
the anomalous electron and muon magnetic moments, which is not given within
the context of the SM. A study of such $g-2$ anomalies in terms of New
Physics and a possible UV complete explanation via vector-like leptons is
performed in \cite{Crivellin:2018qmi}. Also in Ref. \cite{Endo:2019bcj}, it
was shown that the $g-2$ anomalies can be explained using a minimal
supersymmetric standard model assuming a minimal flavor violation in the
lepton sector. %\Antonio{
%Furthermore,
Theories involving extended scalar sector \cite%
{Giudice:2012ms,Falkowski:2018dsl,Crivellin:2018qmi,Allanach:2015gkd,Chen:2016dip,Raby:2017igl,Chiang:2017tai,Chen:2017hir,Davoudiasl:2018fbb,Liu:2018xkx,CarcamoHernandez:2019xkb,Nomura:2019btk,Kawamura:2019rth,Bauer:2019gfk,Han:2018znu,Dutta:2018fge,Badziak:2019gaf,Endo:2019bcj}
as well as vector like leptons \cite{Hiller:2019mou}, heavy $Z^{\prime}$
gauge bosons \cite{CarcamoHernandez:2019ydc,CarcamoHernandez:2019lhv}, and
conformal extended technicolour \cite{Appelquist:2004mn} have been proposed
to explain the $g-2$ anomalies. %}
In this work we will consider a renormalizable theory based on the $%
SU(3)_C\times SU(3)_L\times U(1)_X$ gauge symmetry, supplemented by the
spontaneously broken $U(1)_{L_g}$ global lepton number symmetry and the $S_3
\times Z_2 $ discrete group. We choose $S_3 $ symmetry since it is the
smallest non-Abelian discrete symmetry group having three irreducible
representations (irreps), explicitly, two singlets and one doublet irreps.
This symmetry has been shown to be useful in several extensions of the SM,
for obtaining predictive SM fermion mass matrix textures that successfully
describe the observed SM fermion mass and mixing pattern \cite%
{Gerard:1982mm,Kubo:2003iw,Kubo:2003pd,Kobayashi:2003fh,Chen:2004rr,Mondragon:2007af,Mondragon:2008gm,Bhattacharyya:2010hp, Dong:2011vb,Dias:2012bh,Meloni:2012ci,Canales:2012dr,Canales:2013cga,Ma:2013zca,Kajiyama:2013sza,Hernandez:2013hea, Ma:2014qra,Hernandez:2014vta,Hernandez:2014lpa,Gupta:2014nba,Hernandez:2015dga,Hernandez:2015zeh,Hernandez:2015hrt,Hernandez:2016rbi,CarcamoHernandez:2016pdu,Arbelaez:2016mhg,Gomez-Izquierdo:2017rxi,Cruz:2017add,Ma:2017trv,Espinoza:2018itz,Garces:2018nar,CarcamoHernandez:2018vdj,Gomez-Izquierdo:2018jrx,Pramanick:2019oxb}%
. In the proposed model, the top and exotic quarks get tree level masses
whereas the masses of the bottom, charm and strange quarks as well as the
tau and muon charged lepton masses are produced from a tree level Universal
Seesaw mechanism. The masses for the first generation SM charged fermions
are generated from a one loop level radiative seesaw mechanism mediated by
charged vector like fermions and electrically neutral scalars. The light
active neutrino masses are produced from a one loop level radiative seesaw
mechanism. %\Antonio{
Unlike the 3-3-1 models of Refs. \cite%
{Hernandez:2013hea,Hernandez:2014vta,Hernandez:2014lpa,Hernandez:2015tna,Hernandez:2015cra,Hernandez:2016eod,CarcamoHernandez:2017kra,CarcamoHernandez:2018iel,CarcamoHernandez:2019iwh}%
, where non renormalizable Yukawa interactions are employed for the
implementation of a Froggat Nielsen mechanism to produce the current SM
fermion mass and mixing pattern, after the discrete symmetries are
spontaneously broken, our proposed model is a fully renormalizable theory
with minimal particle content and symmetries where tree level Universal and
a one-loop level radiative seesaw mechanisms are combined to explain the
observed hierarchy of SM fermion masses and fermionic mixing parameters.
Furthermore, unlike Refs. \cite%
{Hernandez:2013hea,Hernandez:2014vta,Hernandez:2014lpa,Hernandez:2015tna,Hernandez:2015cra,Hernandez:2016eod,CarcamoHernandez:2017kra,CarcamoHernandez:2018iel,CarcamoHernandez:2019iwh}
our current work has an explanation for the electron and anomalous magnetic
moments. In our current model, the charged vector-like leptons which mediate
the tree level Universal and one-loop level radiative seesaw mechanism that
generates the SM charged fermion mass hierarchy, make contributions to the
measured values of the muon and electron anomalous magnetic moments, thus
providing a connection of the fermion mass generation mechanism and the $g-2$
anomalies, which is not given the models of Refs. \cite%
{Hernandez:2013hea,Hernandez:2014vta,Hernandez:2014lpa,Hernandez:2015tna,Hernandez:2015cra,Hernandez:2016eod,CarcamoHernandez:2017kra,CarcamoHernandez:2018iel,CarcamoHernandez:2019iwh}%
. %}
Our model is consistent with the low energy SM fermion flavor data and
successfully accommodates the experimental values of the muon and electron
magnetic dipole moments.

The content of this paper goes as follows. The model is introduced in section %
\ref{Model}. The model predictions for the muon and electron anomalous magnetic
moments are discussed in section \ref{gminus2mu}. Section \ref{quarksector} is dedicated to  the quark masses and mixings. Lepton masses
and mixings are analyzed within the model in section \ref{leptonsector}. The generation of
neutrino masses is discussed in section \ref{sec:Neutrino masses and mixings}%
.
Conclusions are given in section \ref{conclusions}.

\section{The model}

\label{Model} We consider a renormalizable extension of the 3-3-1 model with
right handed Majorana neutrinos, where the $SU(3)_C\times SU(3)_L\times
U(1)_X$ gauge symmetry is supplemented by the spontaneously broken $%
U(1)_{L_g}$ global lepton number symmetry and the $S_3 \times Z_2 $ discrete
group, the scalar sector is enlarged by the inclusion of several gauge
singlet scalars and the fermion sector is minimally augmented by the
introduction of heavy electrically charged vector like fermions. Such
electrically charged vector like fermions are assumed to be singlets under
the $SU(3)_L$ gauge symmetry, thus allowing to easily comply with collider
constraints as well as with the constraints arising from electroweak
precision tests. The left and right handed components of such vector like
fermions have the same transformation properties under the different group
factors of the model thus allowing to build mass terms for these fields
invariant under the $SU(3)_C\times SU(3)_L\times U(1)_X\times S_3 \times Z_2 
$ group. The scalar and fermionic content with their assignments under the $%
SU(3)_C\times SU(3)_L\times U(1)_X\times S_3 \times Z_2 $ group are shown in
Tables \ref{tab:scalars} and \ref{tab:fermions}, respectively. The
dimensions of the $SU(3)_C$, $SU(3)_L$ and $S_3 $ representations shown in
Tables \ref{tab:scalars} and \ref{tab:fermions} are specified by the numbers
in boldface. It is worth mentioning that the set of vector like fermions $%
\tilde{T}_{n}$ ($n=1,2$), $B_i$ and $E_i$ ($i=1,2,3$) is the minimum amount
of exotic fermions required to generate the tree level masses via Universal
seesaw mechanism for the bottom, charm and strange quarks as well as the tau
and muon as well as one loop level masses for the first generation SM
charged fermions, i.e., the up, down quarks and the electron. To implement
such tree level Universal and radiative seesaw mechanisms we have introduced
the gauge singlet scalars $\xi_n$ ($n=1,2)$ and $\varphi$. In addition, the
remaining gauge singlet scalars $\sigma_i$ ($i=1,2,3$) are crucial to
generate the Majorana mass terms necessary to radiatively produce the light
active neutrino masses. The vector like fermions mix with the SM charged
fermions lighter than the top quark thus giving rise to a tree level
Universal seesaw mechanism that produces the masses for the bottom, charm
and strange quarks as well as the tau and muon charged lepton masses. The
first generation SM charged fermions, i.e., the up, down quarks and the
electron get their masses from a one loop level radiative seesaw mechanism
mediated by charged vector like fermions and electrically neutral scalars.
In addition, light active neutrino masses are generated from a one loop
level radiative seesaw mechanism mediated by the right handed Majorana
neutrinos and the electrically neutral components of the $SU(3)_L$ scalar
triplet $\chi$. The smallness of the light active neutrino masses is
attributed to a small mass splitting of the $\chi _{1R}$ and $\chi _{1I}$
scalar fields, which originates from the trilinear term $A\left( \chi
^{\dagger }\eta \sigma _{3}+h.c\right)$ of the scalar potential given in
Appendix \ref{potential}. Thus, the trilinear coupling $A$ has to be
sufficiently small to provide a natural explanation for the tiny masses of
the light active neutrinos. In sec. \ref{sec:Neutrino masses and mixings} we
discuss a symmetry-based condition for technically natural smallness of the
parameter $A$. Notice that the $U(1)_{L_g}$ global lepton number symmetry is
spontaneously broken down to a residual discrete $Z_2 ^{(L_g)}$ by the
vacuum expectation value (VEV) of the $U(1)_{L_g}$ charged gauge-singlet
scalars $\sigma _{i}$ ($i=1,2,3$) having a nontrivial $U(1)_{L_g}$ charge,
as indicated by Table \ref{tab:scalars}. The residual discrete $Z_2 ^{(L_g)}$
lepton number symmetry, under which the leptons are charged and the other
particles are neutral, forbids interactions having an odd number of leptons,
thus preventing proton decay. The massless Goldstone boson, i.e., the
Majoron, arising after the spontaneous breaking of the $U(1)_{L_g}$
symmetry, does not cause problems in the model because it is a $SU(3)_L$
scalar singlet.

In addition, our model does not have fermions with exotic electric charges.
Thus, the electric charge in our model is defined as follows: 
\begin{eqnarray}  \label{eq:electric-Charge}
Q=T_3+\beta T_8+X=T_3 -\frac{1}{\sqrt{3}}T_8+X\, .
\end{eqnarray}
Furthermore, the lepton number has a gauge component as well as a
complementary global one, as indicated by the following relation: 
\begin{eqnarray}  \label{eq:Lepton-Number}
L=\frac{4}{\sqrt{3}}T_{8}+L_g,
\end{eqnarray}
being $L_g$ a conserved charge associated with the $U(1)_{L_g}$ global
lepton number symmetry.

In our model the full symmetry $\mathcal{G}$ experiences the following
spontaneous symmetry breaking chain: 
\begin{eqnarray}
&&\mathcal{G}=SU(3)_{C}\times SU(3)_{L}\times U(1)_{X}\times
U(1)_{L_{g}}\times S_{3}\times Z_{2}{\xrightarrow{v_{\chi},v_{\xi},v_{\va },}%
}  \notag \\
&&\hspace{7mm}SU(3)_{C}\times SU(2)_{L}\times U(1)_{L_{g}}{%
\xrightarrow{v_\eta ,v_{\rho}}}  \notag \\
&&\hspace{7mm}SU(3)_{C}\times U(1)_{Q}\times U(1)_{L_{g}}{\xrightarrow{v_\si
,v_{\si _3}}}  \notag \\
&&\hspace{7mm}SU(3)_{C}\times U(1)_{Q},  \label{SB}
\end{eqnarray}%
where the different symmetry breaking scales fulfill the hierarchy: 
\begin{equation}
v_{\chi }\sim v_{\xi }\sim v_{\varphi }\gg v_{\eta },v_{\rho }\gg v_{\sigma
}\sim v_{\sigma _{3}},  \label{VEVhierarchy}
\end{equation}%
with $v_{\eta }^{2}+v_{\rho }^{2}=v^{2}$, $v=246$ GeV. 
% and $v_{\chi }\sim 
%\mathcal{O}(10)$ TeV. 
We assume that the scale $v_{\chi }$ of spontaneous $\ SU(3)_{L}\times
U(1)_{X}$ gauge symmetry breaking is about $10 $ TeV or larger in order to
keep consistency with the collider constraints \cite{Salazar:2015gxa}, the
constraints from the experimental data on $K$, $D $ and $B$-meson mixings 
\cite{Huyen:2012uk} and $B_{s,d}\rightarrow \mu ^{+}\mu ^{-}$, $%
B_{d}\rightarrow K^{\ast }(K)\mu ^{+}\mu ^{-}$ decays \cite%
{CarcamoHernandez:2005ka,Martinez:2008jj,Buras:2013dea,Buras:2014yna,Buras:2012dp}

In principle, the hierarchical VEV pattern (\ref{VEVhierarchy}), being
unprotected by any symmetry, can be affected by large radiative corrections.
The common remedy against this issue is to assume that our model is embedded
into a more fundamental setup with additional symmetries protecting the
hierarchy up to the Planck scale. The well-known examples of such setups are
supersymmetry and warped five-dimensions. Formulation of the corresponding
ultraviolet completion is beyond the scope of the present paper and will be
done elsewhere. %
One can also be concerned about the classical stability of the scalar
potential at the vacuum configuration (\ref{VEVhierarchy}). The latter must
belong to the minimum of the model scalar potential shown in Appendix \ref%
{potential}. This means that the scalar mass squared matrices in the vacuum (%
\ref{VEVhierarchy}) are positively definite.  Having at our
disposal a large number of free parameters in the scalar potential (\ref{V})
it is reasonable to expect that this condition can be easily satisfied in a
wide range of the model parameter space. In section \ref{gminus2mu} we show
that this is true for the benchmark point (\ref{eq:benchmarklambda}) used
for the analysis of $(g-2)_{e,\mu}$.  %In Appendix

%A justification of the above mentioned
%VEV hierarchy is shown in Appendix \ref{apVEVhierarchy}. The following comments about radiative stability of the symmetry breaking scheme (\ref{SB}) and the corresponding VEV hierarchy (\ref{VEVhierarchy}) are in order. It is worth mentioning that that this hierarchy is not protected from large radiative corrections by some symmetry. Therefore, certain tuning of the model parameters are needed to stabilize such hierarchy. The corresponding conditions of vacuum stability can be obtained from the 1-loop Coleman-Weinberg effective potential. Such analysis is beyond the scope of the present paper. However, since in our model the VEV hierarchy  (\ref{VEVhierarchy}) is rather moderate, not exceeding three orders of magnitude, we expect that the quadratic divergences -- dangerous for a strong hierarchy-- can be tamed here by a moderate tuning of the model parameters. At the same time, for the scales above $v_{\chi}$ in (\ref{VEVhierarchy}),  where this is not possible, we assume that our model is embedded into a more fundamental setup with additional symmetries protecting the hierarchy up to the Planck scale. The well-known examples of such setups are supersymmetry and warped five-dimensions. Specifying the details of such ultraviolet completion of our model is left beyond the scope of the present paper and will be done elsewhere.

The $SU(3)_{L}$ triplet scalars $\chi $, $\eta $ and $\rho $ can be expanded
around the minimum as follows: 
\begin{equation}
\chi =%
\begin{pmatrix}
\frac{1}{\sqrt{2}}\left( \chi _{1R}^{0}+i\chi _{1I}^{0}\right) \\ 
\chi _{2}^{-} \\ 
\frac{1}{\sqrt{2}}(v_{\chi }+\xi _{\chi }\pm i\zeta _{\chi })%
\end{pmatrix}%
,\hspace{0.5cm}\eta =%
\begin{pmatrix}
\frac{1}{\sqrt{2}}(v_{\eta }+\xi _{\eta }\pm i\zeta _{\eta }) \\ 
\eta _{2}^{-} \\ 
\frac{1}{\sqrt{2}}\left( \eta _{3R}^{0}+i\eta _{3I}^{0}\right)%
\end{pmatrix}%
,\hspace{0.5cm}\rho =%
\begin{pmatrix}
\rho _{1}^{+} \\ 
\frac{1}{\sqrt{2}}(v_{\rho }+\xi _{\rho }\pm i\zeta _{\rho }) \\ 
\rho _{3}^{+}%
\end{pmatrix}%
,
\end{equation}

The $SU(3)_L$ fermionic antitriplets and triplets are 
\begin{equation}
Q_{nL}=%
\begin{pmatrix}
D_n \\ 
-U_n \\ 
J_n \\ 
\end{pmatrix}%
_L,\hspace{0.5cm}Q_{3L}=%
\begin{pmatrix}
U_3 \\ 
D_3 \\ 
T \\ 
\end{pmatrix}%
_L,\hspace{0.5cm}L_{iL}=%
\begin{pmatrix}
\nu _{i} \\ 
l_{i} \\ 
\nu _{i}^{c} \\ 
\end{pmatrix}%
_L,\hspace{0.5cm}n=1,2,\hspace{0.1cm}i=1,2,3.
\end{equation}
where $l_{1,2,3}=e,\mu,\tau$.

With the field assignment specified in Tables \ref{tab:scalars} and \ref%
{tab:fermions}, the following quark and lepton Yukawa terms arise: 
\begin{eqnarray}
-\mathcal{L}_{Y}^{(q)} &=&y_{T}\overline{Q}_{3L}\chi T_{R}+y_{J}\left( 
\overline{Q}_{L}\chi ^{\ast }J_{R}\right) _{\mathbf{1}}+y_{U}\overline{Q}%
_{3L}\eta U_{3R}+m_{\widetilde{T}}\left( \overline{\widetilde{T}}_{L}%
\widetilde{T}_{R}\right) _{\mathbf{1}}+m_{B}\left( \overline{B}%
_{L}B_{R}\right) _{\mathbf{1}}+m_{B_{3}}\overline{B}_{3L}B_{3R}  \notag \\
&&+x_{T}\left( \overline{Q}_{L}\rho ^{\ast }\widetilde{T}_{R}\right) _{%
\mathbf{1}}+\sum_{n=1}^{2}z_{n}^{(U)}\left( \overline{\widetilde{T}}_{L}\xi
\right) _{\mathbf{1}^{\prime }}U_{nR}+x_{B}\left( \overline{Q}_{L}\eta
^{\ast }B_{R}\right) _{\mathbf{1}}+\sum_{j=1}^{3}z_{j}^{\left( D\right)
}\left( \overline{B}_{L}\xi \right) _{\mathbf{1}^{\prime }}D_{jR}  \notag \\
&&+y_{B}\overline{Q}_{3L}\rho B_{3R}+\sum_{j=1}^{3}x_{j}^{\left( D\right) }%
\overline{B}_{3L}\varphi D_{jR}+H.c,  \label{lyq}
\end{eqnarray}%
\begin{eqnarray}
-\mathcal{L}_{Y}^{(l)} &=&x_{E}\left( \overline{L}_{L}\rho E_{R}\right) _{%
\mathbf{1}}+\sum_{j=1}^{3}z_{j}^{(l)}\left( \overline{E}_{L}\xi \right) _{%
\mathbf{1}^{\prime }}l_{jR}+y_{E}\overline{L}_{3L}\rho
E_{3R}+\sum_{j=1}^{3}x_{j}^{(l)}\overline{E}_{3L}\varphi l_{jR}+m_{E}\left( 
\overline{E}_{L}E_{R}\right) _{\mathbf{1}}+m_{E_{3}}\overline{E}_{3L}E_{3R} 
\notag \\
&&+x_{N}\left( \overline{L}_{L}\chi N_{R}\right) _{\mathbf{\mathbf{1}}}+y_{N}%
\overline{L}_{3L}\chi N_{3R}+h_{1N}\left( N_{R}\overline{N_{R}^{C}}\right) _{%
\mathbf{\mathbf{2}}}\sigma +h_{2N}\left( N_{R}\overline{N_{3R}^{C}}\sigma
\right) _{\mathbf{1}^{\prime }}  \notag \\
&&+h_{3N}\left( N_{R}\overline{N_{R}^{C}}\right) _{\mathbf{\mathbf{1}}%
}\sigma _{3}+h_{4N}N_{3R}\overline{N_{3R}^{C}}\sigma _{3}+H.c.  \label{lyl}
\end{eqnarray}%
We consider the following VEV configurations for the $S_{3}$ doublets: 
\begin{equation}
\left\langle \xi \right\rangle =v_{\xi }\left( 1,0\right) ,\hspace{0.5cm}%
\left\langle \sigma \right\rangle =\left( v_{\sigma _{1}},v_{\sigma
_{2}}\right) ,  \label{vevxisigma}
\end{equation}%
which are consistent with the scalar potential minimization equations for a
large region of parameter space \cite%
{Kubo:2004ps,Hernandez:2014vta,Hernandez:2015dga}.

\begin{table}[th]
\begin{tabular}{|c|c|c|c|c|c|c|c|}
\hline
& $\chi $ & $\eta $ & $\rho $ & $\xi $ & $\varphi $ & $\sigma $ & $\sigma
_{3}$ \\ \hline
$SU(3)_{C}$ & $1^{\prime }$ & $\mathbf{1}$ & $\mathbf{1}$ & $\mathbf{1}$ & $%
\mathbf{1}$ & $\mathbf{1}$ & $\mathbf{1}$ \\ \hline
$SU(3)_{L}$ & $\mathbf{3}$ & $\mathbf{3}$ & $\mathbf{3}$ & $\mathbf{1}$ & $%
\mathbf{1}$ & $\mathbf{1}$ & $\mathbf{1}$ \\ \hline
$U(1)_{X}$ & $-\frac{1}{3}$ & $-\frac{1}{3}$ & $\frac{2}{3}$ & $0$ & $0$ & $%
0 $ & $0$ \\ \hline
$U(1)_{L_{g}}$ & $\frac{4}{3}$ & $-\frac{2}{3}$ & $-\frac{2}{3}$ & $0$ & $0$
& $2$ & $2$ \\ \hline
$S_{3}$ & $\mathbf{1}$ & $\mathbf{1}$ & $\mathbf{1}$ & $\mathbf{2}$ & $%
\mathbf{1}^{\prime }$ & $\mathbf{2}$ & $\mathbf{1}$ \\ \hline
$Z_{2}$ & $-1$ & $1$ & $1$ & $-1$ & $-1$ & $1$ & $1$ \\ \hline
\end{tabular}%
\caption{Scalar assignments under $SU(3)_{C}\times SU(3)_{L}\times
U(1)_{X}\times U(1)_{L_{g}}\times S_{3}\times Z_{2}$.}
\label{tab:scalars}
\end{table}

\begin{table}[th]
\begin{tabular}{|c|c|c|c|c|c|c|c|c|c|c|c|c|c|c|c|c|c|c|c|c|c|c|}
\hline
& $Q_{L}$ & $Q_{3L}$ & $U_{nR}$ & $U_{3R}$ & $D_{iR}$ & $T_{R}$ & $J_{R}$ & $%
\widetilde{T}_{L}$ & $\widetilde{T}_{R}$ & $B_{L}$ & $B_{R}$ & $B_{3L}$ & $%
B_{3R}$ & $L_{L}$ & $L_{3L}$ & $l_{iR}$ & $E_{L}$ & $E_{R}$ & $E_{3L}$ & $%
E_{3R}$ & $N_{R}$ & $N_{3R}$ \\ \hline
$SU(3)_{C}$ & $\mathbf{3}$ & $\mathbf{3}$ & $\mathbf{3}$ & $\mathbf{3}$ & $%
\mathbf{3}$ & $\mathbf{3}$ & $\mathbf{3}$ & $\mathbf{3}$ & $\mathbf{3}$ & $%
\mathbf{3}$ & $\mathbf{3}$ & $\mathbf{3}$ & $\mathbf{3}$ & $\mathbf{1}$ & $%
\mathbf{1}$ & $\mathbf{1}$ & $\mathbf{1}$ & $\mathbf{1}$ & $\mathbf{1}$ & $%
\mathbf{1}$ & $\mathbf{1}$ & $\mathbf{1}$ \\ \hline
$SU(3)_{L}$ & $\overline{\mathbf{3}}$ & $\mathbf{3}$ & $\mathbf{1}$ & $%
\mathbf{1}$ & $\mathbf{1}$ & $\mathbf{1}$ & $\mathbf{1}$ & $\mathbf{1}$ & $%
\mathbf{1}$ & $\mathbf{1}$ & $\mathbf{1}$ & $\mathbf{1}$ & $\mathbf{1}$ & $%
\mathbf{3}$ & $\mathbf{3}$ & $\mathbf{1}$ & $\mathbf{1}$ & $\mathbf{1}$ & $%
\mathbf{1}$ & $\mathbf{1}$ & $\mathbf{1}$ & $\mathbf{1}$ \\ \hline
$U(1)_{X}$ & $0$ & $\frac{1}{3}$ & $\frac{2}{3}$ & $\frac{2}{3}$ & $-\frac{1%
}{3}$ & $\frac{2}{3}$ & $-\frac{1}{3}$ & $\frac{2}{3}$ & $\frac{2}{3}$ & $-%
\frac{1}{3}$ & $-\frac{1}{3}$ & $-\frac{1}{3}$ & $-\frac{1}{3}$ & $-\frac{1}{%
3}$ & $-\frac{1}{3}$ & $-\frac{1}{3}$ & $-1$ & $-1$ & $-1$ & $-1$ & $0$ & $0$
\\ \hline
$U(1)_{L_{g}}$ & $\frac{2}{3}$ & $-\frac{2}{3}$ & $0$ & $0$ & $0$ & $-2$ & $%
2 $ & $0$ & $0$ & $0$ & $0$ & $0$ & $0$ & $\frac{1}{3}$ & $\frac{1}{3}$ & $1$
& $1$ & $1$ & $1$ & $1$ & $-1$ & $-1$ \\ \hline
$S_{3}$ & $\mathbf{2}$ & $\mathbf{1}$ & $\mathbf{1}^{\prime }$ & $\mathbf{1}$
& $\mathbf{1}^{\prime }$ & $1^{\prime }$ & $\mathbf{2}$ & $\mathbf{2}$ & $%
\mathbf{2}$ & $\mathbf{2}$ & $\mathbf{2}$ & $\mathbf{1}$ & $\mathbf{1}$ & $%
\mathbf{2}$ & $\mathbf{1}$ & $\mathbf{1}^{\prime }$ & $\mathbf{2}$ & $%
\mathbf{2}$ & $\mathbf{1}$ & $\mathbf{1}$ & $\mathbf{2}$ & $1^{\prime }$ \\ 
\hline
$Z_{2}$ & $1$ & $1$ & $-1$ & $1$ & $-1$ & $-1$ & $1$ & $1$ & $1$ & $1$ & $1$
& $1$ & $1$ & $1$ & $1$ & $-1$ & $1$ & $1$ & $1$ & $1$ & $1$ & $-1$ \\ \hline
\end{tabular}%
\caption{Fermion assignments under $SU(3)_{C}\times SU(3)_{L}\times
U(1)_{X}\times U(1)_{L_{g}}\times S_{3}\times Z_{2}$. Here $B_{L,R}=\left(
B_{1(L,R)},B_{2(L,R)}\right) $, $E_{L,R}=\left( E_{1(L,R)},E_{2(L,R)}\right) 
$, $L_{L}=\left( L_{1(L)},L_{2(L)}\right) $, $N_{R}=\left(
N_{1(R)},N_{2(R)}\right) $, $n=1,2$ and $i=1,2,3$.}
\label{tab:fermions}
\end{table}

\section{Muon and electron anomalous magnetic moments}

\label{gminus2mu} The current experimental data on the anomalous dipole
magnetic moments of electron and muon $a_{e,\mu }=(g_{e,\mu }-2)/2$ show
significant deviation from their SM values 
\begin{eqnarray}
\Delta a_{\mu } &=&a_{\mu }^{\mathrm{exp}}-a_{\mu }^{\mathrm{SM}}=\left(
26.1\pm 8\right) \times 10^{-10}\hspace{17mm}\mbox{%
\cite{Hagiwara:2011af,Davier:2017zfy,Nomura:2018lsx,Nomura:2018vfz,Blum:2018mom,Keshavarzi:2018mgv,Aoyama:2020ynm}}\label{eq:a-mu} \\
\Delta a_{e} &=&a_{e}^{\mathrm{exp}}-a_{e}^{\mathrm{SM}}=(-0.88\pm
0.36)\times 10^{-12}\hspace{10mm}\mbox{\cite{Parker:2018vye}}  \label{eq:a-e}
\end{eqnarray}%

\begin{figure}[h]
%\resizebox{18cm}{25cm}{\hspace{-2cm}\includegraphics{Diagramsgminus2c}} 
%\vspace{-18cm}
\includegraphics[width=0.8\textwidth]{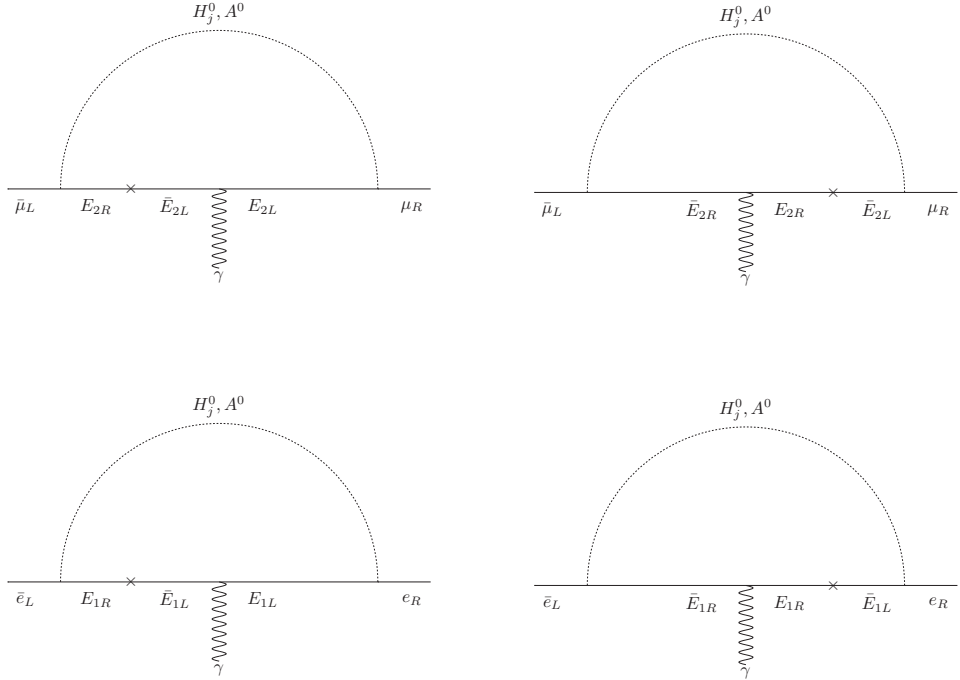}\vspace{-5cm}
\caption{Leading Loop Feynman diagrams contributing to the muon and electron
anomalous magnetic moments. Here $E_{1,2}$, are components of the $S_{3}$%
-doublet and $j=1,2,3,4$.}
\label{Loopdiagramsgminus2}
\end{figure}

Here we analyze predictions of our model for these observables. The leading
contributions to $\Delta a_{e,\mu }$ arising in the model are shown in Fig.~%
\ref{Loopdiagramsgminus2}. The diagrams involve the electrically neutral
physical CP even $H^{0}_{i}$ ($i=1,2,3,4$)\ and CP odd $A^{0}$ scalar as
well as heavy exotic charged $E_{L,R}$ leptons. The physical CP even scalars
arise from the combinations of $\xi _{\rho }$, $\xi _{1R}$, $\xi _{2R}$, $%
\varphi _{R}$ whereas the CP odd scalar corresponds to $\zeta _{\rho }$. By $%
\xi _{1R,2R}$\ we denote real part of the two components of the scalar $%
S_{3} $-doublet gauge singlet $\xi $. Similary, the real part of the scalar $%
S_{3}$-singlet gauge singlet $\varphi $ is denoted by $\varphi _{R}$.
Analogously, $E_{1,2}$ are two components of the leptonic $S_{3}$-doublet
gauge singlet $E_{L,R}$. The fields $\xi _{\rho }$ and $\zeta _{\rho }$ are
contained in the $SU(3)_{L}$ scalar triplet $\rho $, which interacts with $l$%
, the second component of the leptonic triplet $L_{L}$. It is worth
mentioning that, in view of the large amount of parametric freedom of the
model scalar potential in Eq. (\ref{V}), we are restricting to a particular
benchmark scenario were the $SU\left( 3\right) _{L}$ scalar triplet $\rho $
and the gauge singlet scalars $\xi $ and $\varphi $ do not feature mixings
with the remaining scalar fields $\eta $, $\sigma $ and $\sigma _{3}$. Such
benchmark scenario is consistent with the decoupling limit where the CP even
neutral component of the $SU\left( 3\right) _{L}$ scalar triplet $\eta $
mostly corresponds to the $126$ GeV SM like Higgs boson. Another motivation
for such benchmark scenario is the fact that the VEV of the $SU\left(
3\right) _{L}$ scalar triplet $\chi $ is much larger than the VEV of the $%
SU\left( 3\right) _{L}$ scalar triplet $\rho $, thus allowing to neglect the
mixing angles between those fields since they are suppressed by the ratios
of their VEVs, as follows from the method of recursive expansion of Ref. 
\cite{Grimus:2000vj}. Due to the same argument, the mixing angles of the $%
\rho $, $\xi $ and $\varphi $ scalar fields with the gauge singlet scalars $%
\sigma $ and $\sigma _{3}$ can be neglected as well. 

In this framework, the scalar potential terms contributing to the Yukawa
couplings of the fermions $E_1$ and $E_2$ with the scalar fields are shown
in Appendix \ref{appotentialg2} . Let us note the the following
peculiar pattern of mixing in the scalar sector. The fields $\rho$, $\xi $
and $\varphi $ do not mix with $\eta $, $\sigma $ and $\sigma_{3}$, while $%
\varphi$ mix with $\zeta_{\rho}$ through the the complex parameter $\kappa$
in the scalar potential \eqref{V}. In view of this we find that the scalar
mass matrix in the basis $\xi_{\rho}$, $\xi_{1R}$, $\xi_{2R}$, $\varphi$,
and $\zeta_{\rho}$ has the form: 
\begin{align}  \label{eq:M2-S}
\mathbf{M}^{2}&= 
\begin{pmatrix}
m^{2}_{11} & m^{2}_{12} & 0 & m^{2}_{14} & 0 \\ 
m^{2}_{21} & m^{2}_{22} & m^{2}_{23} & m^{2}_{24} & 0 \\ 
0 & m_{32} & m_{33} & m_{34} & 0 \\ 
m^{2}_{41} & m^{2}_{42} & m^{2}_{43} & m^{2}_{44} & m^{2}_{45} \\ 
0 & 0 & 0 & m^{2}_{54} & m^{2}_{55} \\ 
&  &  &  & 
\end{pmatrix}%
\end{align}
with the matrix elements $m^{2}_{ij}$ given in Appendix \ref{appotentialg2}
. Once the basis is changed by a rotation matrix $R$, the physical scalar
field masses $m_{H^{0}_{1}}$, $m_{H^{0}_{2}}$, $m_{H^{0}_{3}}$, $%
m_{H^{0}_{5}}$ and $m_{H^{0}_{A}}$ can be found numerically. 

Thus, in our model the muon and electron anomalous magnetic moments are
given by: 
\begin{equation}
\Delta a_{\mu ,e}=\dsum\limits_{i=1}^{4}\dsum\limits_{\Phi
=H^{0}_{i},A^{0}}\Delta a_{\mu ,e}(\Phi ),
\end{equation}
 The analytical form for the neutral scalar contribution at one loop to $%
\Delta a_{\mu ,e}$ can be found in \cite%
{Diaz:2002uk,Kelso:2014qka,Lindner:2016bgg,Kowalska:2017iqv}. Using these
results we write the contributions of the neutral scalars $\Phi =H^{0},A^{0}$
as follows:

{%
\begin{equation}
\Delta a_{\mu }=w_{\mu }^{2}\frac{m_{\mu }^{2}}{8\pi ^{2}}\left\{
\dsum\limits_{i=1}^{4}\left( R^{T}\right) _{1i}\left( R^{T}\right) _{2i}%
\frac{G_{S}^{\left( l\right) }\left( m_{E_{2}},m_{H_{i}^{0}}\right) }{%
m_{H_{i}^{0}}^{2}}+\left( R^{T}\right) _{55}\left( R^{T}\right) _{25}\frac{%
G_{P}^{\left( l\right) }\left( m_{E_{2}},m_{A^{0}}\right) }{m_{A^{0}}^{2}}%
\right\}  \label{Deltaamu}
\end{equation}%
}

{%
\begin{equation}
\Delta a_{e}=w_{e}^{2}\frac{m_{\mu }^{2}}{8\pi ^{2}}\left\{
\dsum\limits_{i=1}^{4}\left( R^{T}\right) _{1i}\left( R^{T}\right) _{3i}%
\frac{G_{S}^{\left( l\right) }\left( m_{E_{1}},m_{H_{i}^{0}}\right) }{%
m_{H_{i}^{0}}^{2}}+\left( R^{T}\right) _{55}\left( R^{T}\right) _{35}\frac{%
G_{P}^{\left( l\right) }\left( m_{E_{1}},m_{A^{0}}\right) }{m_{A^{0}}^{2}}%
\right\}  \label{Deltaae}
\end{equation}%
} %
where the loop function is given by: 
\begin{equation}
G_{S,P}^{(l)}(m_{E}, m_{\Phi} )=\int_{0}^{1}dx\frac{x^{2}(1-x\pm \epsilon
_{lE})}{(1-x)(1-x\lambda _{l\Phi }^{2})+x\epsilon _{lE}^{2}\lambda _{l\Phi
}^{2}},\hspace{1cm}\Phi =H^{0},A^{0}  \label{Gloop}
\end{equation}
with $l=e,\mu $ and $\lambda _{l\Phi }=m_{l}/m_{\Phi }$, $\epsilon
_{eE}=m_{E_{1}}/m_{e}$, $\epsilon _{\mu E}=m_{E_{2}}/m_{\mu}$. 

Besides that, the plus and minus signs for the loop function $G_{S,P}(\Phi )$
of Eq. (\ref{Gloop}) stands for the scalar (CP-even) and pseudoscalar
(CP-odd) contributions, respectively.\ The quantities $w_{l}$ ($l=e,\mu $)\
are the Yukawa couplings for the interaction $w_{l}\overline{E}l\,\Phi $.

\begin{figure}[h]
\centering
\subcaptionbox{Correlation plot between the mass of the exotic fermion $E_2$
and the value of $\Delta
a_\mu$.}{\includegraphics[width=.49\textwidth]{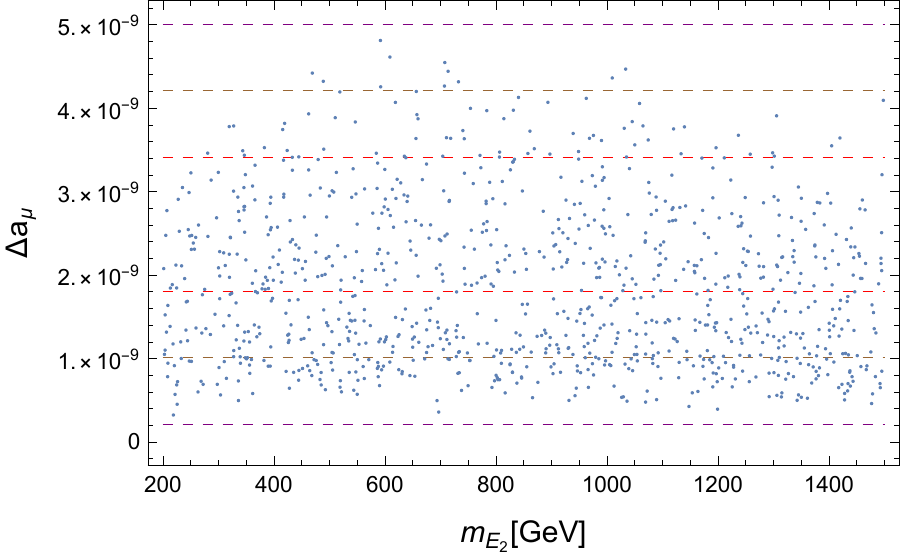}}\hfill 
\subcaptionbox{Correlation plot between the mass of the exotic fermion $E_1$
and the value of $\Delta
a_e$.}{\includegraphics[width=.49\textwidth]{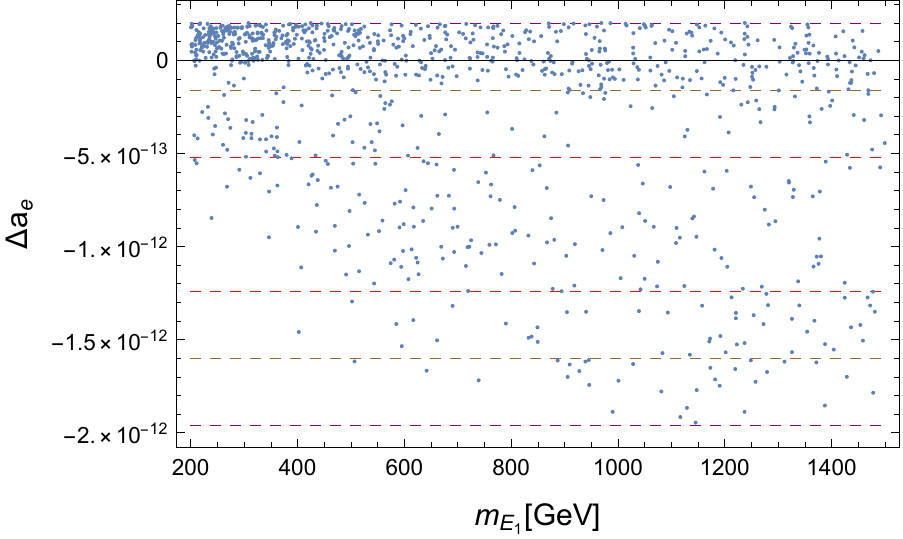}}\hfill
\caption{Correlation plots of the $\Delta a_{\protect\mu, e}$ and the mass
of the exotic fermion $m_{E_{2,1}}$ respectively at $1\protect\sigma$ (red), 
$2\protect\sigma$ (brown) and $3\protect\sigma$ (purple).}
\label{gminus}
\end{figure}

The experimental values of the muon and electron anomalous magnetic moments
shown in Eqs. \eqref{eq:a-mu} and \eqref{eq:a-e} can be successfully
reproduced at $2\sigma$ level for the following benchmark point: 
\begin{align}
v_{\eta} &= v_{\rho} \approx 174\,\text{GeV} & v_{\chi} &\approx 2851\,\text{%
GeV} & v_{\xi} &\approx 1414\,\text{GeV}  \notag \\
v_{\phi} &\approx 6992\,\text{GeV} & \kappa_r &\approx -0.630 & \kappa_i
&\approx -0.614  \notag \\
\lambda_{2}&= \lambda_{10} \approx 7.251 & \lambda_{12} &= \lambda_{13}
\approx 0.310 & \lambda_{18} &= \lambda_{34} \approx -0.264  \notag \\
& & \lambda_{35} &= \lambda_{38} \approx -0.229 & &
\label{eq:benchmarklambda}
\end{align}
The scalar and charged exotic leptons masses along with the
Yukawa couplings are  %\begin{align}
\begin{align}
m_{H^0_1} &\approx 5786\,\text{GeV} & m_{H^0_2} &\approx 5338\,\text{GeV} & 
m_{H^0_3} &\approx 2750\,\text{GeV}  \notag \\
m_{H^0_4} &\approx 2498\,\text{GeV} & m_{A^0} &\approx 1100 \,\text{GeV} & 
m_{E_1} &\approx 611\,\text{GeV}  \notag \\
m_{E_2} &\approx 1368\,\text{GeV} & w_{\mu} &\approx 0.228 & w_{e} &\approx
1.719  \label{eq:benchmarkmass}
\end{align}
Note that this benchmark point locates in the domain of the
model parameter space corresponding to the minimum of the scalar potential
due to the fact that all the scalar masses are real (see also Appendix \ref%
{appotentialg2}). In this benchmark point the muon and electron $(g-2)$%
-experimental anomalies have the values % 
\begin{align}  \label{eq:benchmarkg-1-2}
\Delta a_{\mu } &= 2.68714 \times 10^{-9} \\
\Delta a_{e} &= -8.64531 \times 10^{-13}  \label{eq:benchmarkg-2}
\end{align}
The opposite signs of these quantities is due to the pseudo
scalar $A^{0}$ contributions to the loops in Fig.~\ref{Loopdiagramsgminus2}
leading to the minus sign in the term $-\epsilon_{lE}$ of the loop function (%
\ref{Gloop}). Note that $E_{2}$ and $E_{1}$ contribute separately to the
muon and electron $(g-2)$, respectively, without any cross-contributions.
Thus, selecting appropriate values for the exotic lepton masses $m_{E_{1,2}}$
we can accommodate the experimental sign difference (\ref{eq:benchmarkg-1-2}%
), (\ref{eq:benchmarkg-2}). The fact that $m_{e}\ll m_{\mu}$ makes the
required sign difference valid in a wide range of the model parameter space.
To show this, let us vary the model parameters within 15\% around the
benchmark point \eqref{eq:benchmarklambda} and the charged exotic lepton
masses in a range from $200$~GeV to $1200$~GeV. The resulting $\Delta
a_{\mu,e}-m_{E_{2,1}}$ scatter plots are shown in Figure~\ref{gminus}. As
can be seen, the model indeed can explain the experimental values of muon
and electron anomalous magnetic moments simultaneously in a wide range of
its parameter space. 

\section{Quark masses and mixings}

\label{quarksector}

\begin{figure}[h]
\resizebox{16cm}{22cm}{\vspace{-2cm}\includegraphics{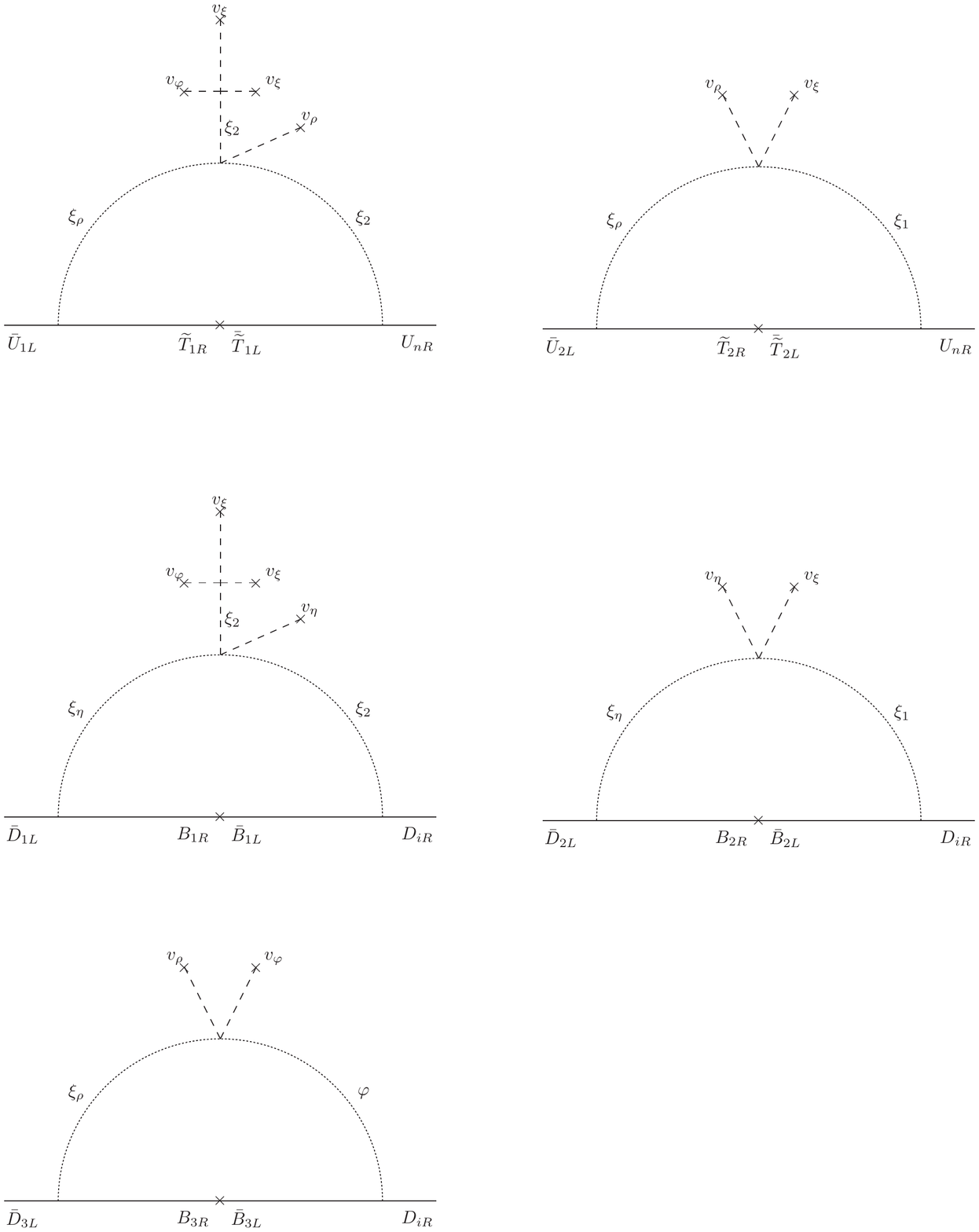}}%
\vspace{0cm}
\caption{One-loop Feynman diagrams contributing to the entries of the SM
quark mass matrices. Here, $n=1,2$ and $i=1,2,3$.}
\label{Loopdiagramsq}
\end{figure}

From the quark Yukawa interactions in Eq. (\ref{lyq}), we find that the
up-type mass matrix in the basis $(\overline{u}_{1L},\overline{u}_{2L},%
\overline{u}_{3L},\overline{T}_{L},\overline{\widetilde{T}}_{1L},\overline{%
\widetilde{T}}_{2L})$ versus $(u_{1R},u_{2R},u_{3R},T_{R},\widetilde{T}_{1R},%
\widetilde{T}_{2R})$ is given by: 
\begin{eqnarray}
M_{U} &=&\left( 
\begin{array}{cccc}
\Delta _{U} & 0_{2\times 1} & 0_{2\times 1} & A_{U} \\ 
0_{1\times 2} & m_{t} & 0 & 0_{1\times 2} \\ 
0_{1\times 2} & 0 & m_{T} & 0_{1\times 2} \\ 
B_{U} & 0_{2\times 1} & 0_{2\times 1} & \widetilde{M}_{T}%
\end{array}%
\right) ,\hspace{0.5cm}A_{U}=x_{T}\left( 
\begin{array}{cc}
1 & 0 \\ 
0 & 1%
\end{array}%
\right) \frac{v_{\rho }}{\sqrt{2}},\hspace{0.5cm}B_{U}=\left( 
\begin{array}{cc}
0 & 0 \\ 
z_{1}^{(U)} & z_{2}^{(U)}%
\end{array}%
\right) v_{\xi },  \notag \\
m_{t} &=&y_{U}\frac{v_{\eta }}{\sqrt{2}}=a_{3}^{(U)}\frac{v}{\sqrt{2}},%
\hspace{0.2cm}\widetilde{M}_{T}=m_{\widetilde{T}}\left( 
\begin{array}{cc}
1 & 0 \\ 
0 & 1%
\end{array}%
\right) ,\hspace{0.2cm}\Delta _{U}=\left( 
\begin{array}{cc}
\varepsilon _{11}^{(U)} & \varepsilon _{12}^{(U)} \\ 
\varepsilon _{21}^{(U)} & \varepsilon _{22}^{(U)}%
\end{array}%
\right) \frac{v_{\rho }}{\sqrt{2}},  \notag \\
\varepsilon _{1n}^{(U)} &=&\frac{1}{16\pi ^{2}}\frac{\lambda _{\rho
^{\dagger }\rho \xi ^{2}}\lambda _{\xi ^{3}\varphi
}x_{T}z_{n}^{(U)}v_{\varphi }v_{\xi }^{2}}{m_{T}m_{\xi _{2}}^{2}}\left[
C_{0}\left( \frac{m_{\xi _{\eta }}}{m_{B}},\frac{m_{\func{Re}\xi _{2}}}{m_{B}%
}\right) -C_{0}\left( \frac{m_{\zeta _{\eta }}}{m_{B}},\frac{m_{\func{Im}\xi
_{2}}}{m_{B}}\right) \right] ,\hspace{0.2cm}n=1,2,  \notag \\
\varepsilon _{2n}^{(U)} &=&\frac{1}{16\pi ^{2}}\frac{\lambda _{\rho
^{\dagger }\rho \xi ^{2}}x_{T}z_{n}^{(U)}v_{\xi }}{m_{T}}\left[ C_{0}\left( 
\frac{m_{\xi _{\eta }}}{m_{B}},\frac{m_{\func{Re}\xi _{2}}}{m_{B}}\right)
-C_{0}\left( \frac{m_{\zeta _{\eta }}}{m_{B}},\frac{m_{\func{Im}\xi _{2}}}{%
m_{B}}\right) \right] ,  \label{MU}
\end{eqnarray}

whereas the down type quark mass matrix written in the basis \newline
$(\overline{d}_{1L},\overline{d}_{2L},\overline{d}_{3L},\overline{J}_{1L},%
\overline{J}_{2L},\overline{B}_{1L},\overline{B}_{2L},\overline{B}_{3L})$-$%
(d_{1R},d_{2R},d_{3R},J_{1R},J_{2R},B_{1R},B_{2R},B_{3R})$ takes the form: 
\begin{eqnarray}
M_{D} &=&\left( 
\begin{array}{ccc}
\Delta _{D} & 0_{3\times 2} & A_{D} \\ 
0_{2\times 3} & M_{J} & 0_{2\times 3} \\ 
B_{D} & 0_{3\times 2} & M_{B}%
\end{array}%
\right) ,\hspace{0.1cm}A_{D}=\left( 
\begin{array}{ccc}
x_{B}\frac{v_{\eta }}{\sqrt{2}} & 0 & 0 \\ 
0 & x_{B}\frac{v_{\eta }}{\sqrt{2}} & 0 \\ 
0 & 0 & y_{B}\frac{v_{\rho }}{\sqrt{2}}%
\end{array}%
\right) ,\hspace{0.1cm}B_{D}=\left( 
\begin{array}{ccc}
0 & 0 & 0 \\ 
z_{1}^{(D)}v_{\xi } & z_{2}^{(D)}v_{\xi } & z_{3}^{(D)}v_{\xi } \\ 
x_{1}^{(D)}v_{\varphi } & x_{2}^{(D)}v_{\varphi } & x_{3}^{(D)}v_{\varphi }%
\end{array}%
\right) ,  \notag \\
M_{J} &=&y^{\left( J\right) }\frac{v_{\chi }}{\sqrt{2}}\left( 
\begin{array}{cc}
1 & 0 \\ 
0 & 1%
\end{array}%
\right) ,\hspace{1cm}M_{B}=\left( 
\begin{array}{ccc}
m_{B} & 0 & 0 \\ 
0 & m_{B} & 0 \\ 
0 & 0 & m_{B_{3}}%
\end{array}%
\right) ,\hspace{1cm}\Delta _{D}=\left( 
\begin{array}{ccc}
\varepsilon _{11}^{(D)} & \varepsilon _{12}^{(D)} & \varepsilon _{13}^{(D)}
\\ 
\varepsilon _{21}^{(D)} & \varepsilon _{22}^{(D)} & \varepsilon _{23}^{(D)}
\\ 
\varepsilon _{31}^{(D)} & \varepsilon _{32}^{(D)} & \varepsilon _{33}^{(D)}%
\end{array}%
\right) \frac{v_{\rho }}{\sqrt{2}},  \notag \\
\varepsilon _{1i}^{(D)} &=&\frac{1}{16\pi ^{2}}\frac{\lambda _{\rho
^{\dagger }\rho \xi ^{2}}\lambda _{\xi ^{3}\varphi
}x_{B}z_{i}^{(D)}v_{\varphi }v_{\xi }^{2}v_{\eta }}{m_{B}m_{\xi
_{2}}^{2}v_{\rho }}C_{0}\left( \frac{m_{\xi _{\eta }}}{m_{B}},\frac{m_{\func{%
Re}\xi _{2}}}{m_{B}}\right) ,  \notag \\
\varepsilon _{2i}^{(D)} &=&\frac{1}{16\pi ^{2}}\frac{\lambda _{\rho
^{\dagger }\rho \xi ^{2}}x_{B}z_{i}^{(D)}v_{\xi }v_{\eta }}{m_{B}v_{\rho }}%
C_{0}\left( \frac{m_{\xi _{\eta }}}{m_{B}},\frac{m_{\func{Re}\xi _{2}}}{m_{B}%
}\right) ,  \notag \\
\varepsilon _{3i}^{(D)} &=&\frac{1}{16\pi ^{2}}\frac{\lambda _{\rho
^{\dagger }\rho \varphi ^{2}}x_{B}z_{i}^{(D)}v_{\varphi }}{m_{B}}C_{0}\left( 
\frac{m_{\xi _{\rho }}}{m_{B}},\frac{m_{\func{Re}\varphi }}{m_{B}}\right) ,%
\hspace{0.2cm}i=1,2,3,  \label{MD}
\end{eqnarray}

where as seen from Eqs. (\ref{MU}) and (\ref{MD}), the $\Delta _U $ and $%
\Delta_D $ submatrices are generated at one loop level. The one loop level
Feynman diagrams generating the $\Delta _U $ and $\Delta _D $ submatrices
are shown in Figure \ref{Loopdiagramsq}. In addition, the following function
has been introduced: 
\begin{equation}
C_{0}\left( \widehat{m}_1 ,\widehat{m}_2 \right) =\frac{1}{\left( 1-\widehat{%
m}_1 ^2 \right) \left( 1-\widehat{m}_2 ^2 \right) \left( \widehat{m}_1 ^2 -%
\widehat{m}_2 ^2 \right) }\left\{ \widehat{m}_1 ^2 \widehat{m}_2 ^2 \ln
\left( \frac{\widehat{m}_1 ^2 }{\widehat{m}_2 ^2 }\right) -\widehat{m}_1 ^2
\ln \widehat{m}_1 ^2 +\widehat{m}_2 ^2 \ln \widehat{m}_2 ^2 \right\} .
\label{loopfunction}
\end{equation}

%As shown by
%\Sergey{
As seen from Eqs. (\ref{MU}) and (\ref{MD}), %the very heavy vector like
the exotic heavy vector like quarks mix with the SM quarks lighter than top
quark. %Such very heavy
The masses of these exotic %vector like
quarks are assumed to be %have masses
much larger than the $SU(3) _L\times U(1) _X$ symmetry breaking scale. 
%Thus, in view of the above, a
%Therefore,
As a result, charm, bottom and strange quarks acquire their masses from the
tree-level Universal seesaw mechanism, while 
% produces the charm, bottom and strange quark masses.
the masses of the up and down quarks are generated by the one-loop radiative
seesaw mechanism. %Consequently,
Thus, for the SM quarks we obtain the following mass matrices: 
\begin{equation}
\widetilde{M}_U =\left( 
\begin{array}{cc}
\Delta _U +A_U M_{\widetilde{T}}^{-1}B_U & 0_{2\times 1} \\ 
0_{1\times 2} & m_{t}%
\end{array}%
\right) =\left( 
\begin{array}{ccc}
\varepsilon _{11}^{(U) }\frac{v_\rho}{\sqrt{2}} & \varepsilon _{12}^{(U) }%
\frac{v_\rho}{\sqrt{2}} & 0 \\ 
\varepsilon _{21}^{(U) }\frac{v_\rho}{\sqrt{2}}+\frac{x_{T}z_1 ^{(U) }v_\xi
v_\rho}{\sqrt{2}m_{\widetilde{T}}} & \varepsilon _{22}^{(U) }\frac{v_\rho}{%
\sqrt{2}}+\frac{x_{T}z_2 ^{(U) }v_\xi v_\rho}{\sqrt{2}m_{\widetilde{T}}} & 0
\\ 
0 & 0 & m_{t}%
\end{array}%
\right),
\end{equation}%
\begin{eqnarray}
\widetilde{M}_D &=&\Delta _D +A_D M_B ^{-1}B_D  \notag \\
&=&\left( 
\begin{array}{ccc}
\varepsilon _{11}^{(D) }\frac{v_\rho}{\sqrt{2}} & \varepsilon _{12}^{(D) }%
\frac{v_\rho}{\sqrt{2}} & \varepsilon _{13}^{(D) }\frac{v_\rho}{\sqrt{2}} \\ 
\varepsilon _{21}^{(D) }\frac{v_\rho}{\sqrt{2}}+x_B z_1 ^{(D) }\frac{v_\xi
v_\eta }{\sqrt{2}m_B } & \varepsilon _{22}^{(D) }\frac{v_\rho}{\sqrt{2}}+x_B
z_2 ^{(D) }\frac{v_\xi v_\eta }{\sqrt{2}m_B } & \varepsilon _{23}^{(D) }%
\frac{v_\rho}{\sqrt{2}}+x_B z_3 ^{(D) }\frac{v_\xi v_\eta }{\sqrt{2}m_B } \\ 
\varepsilon _{31}^{(D) }\frac{v_\rho}{\sqrt{2}}+y_B x_1 ^{(D) }\frac{%
v_{\varphi }v_\rho}{\sqrt{2}m_{B_3 }} & \varepsilon _{32}^{(D) }\frac{v_\rho%
}{\sqrt{2}}+y_B x_2 ^{(D) }\frac{v_{\varphi }v_\rho}{\sqrt{2}m_{B_3 }} & 
\varepsilon _{33}^{(D) }\frac{v_\rho}{\sqrt{2}}+y_B x_3 ^{(D) }\frac{%
v_{\varphi }v_\rho}{\sqrt{2}m_{B_3 }}%
\end{array}%
\right).
\end{eqnarray}

These mass matrices contain several model parameters. While free, they still
satisfy certain conditions in our model. In fact, vev's $v_{\xi,\eta,\rho}$
obey the inequality (\ref{VEVhierarchy}) expressing the hierarchy of
symmetry breaking in our model. The $\varepsilon_{ij}^{U,D}$ parameters,
defined in Eqs.~(\ref{MU}) and (\ref{MD}), contain typical loop suppression
and specific dependence on the vev's, exotic masses, the Yukawas and a
quartic coupling. We require the latter to satisfy the perturbativity
condition. With this in mind we can speak about natural values of the matrix
elements corresponding to the values of the model parameters in a range not
involving an ad hoc hierarchy of the dimensionless couplings. 
%natural values
Let us show that within this natural range the model accommodates the
observable values of the SM quark masses and mixings. To this end we
consider a particular natural benchmark scenario consistent with the
above-mentioned conditions. We choose: 
\begin{eqnarray}  \label{eq:quark-bmp-2}
v_\xi &=&\lambda ^{4}\frac{vm_{\widetilde{T}}}{v_\rho}=\lambda ^5\frac{vm_B 
}{v_\eta },\hspace{0.5cm}v_{\varphi }=\lambda ^3 \frac{vm_{B_3 }}{v_\rho},%
\hspace{0.5cm}\varepsilon_{nm}^{(U) }=b_{nm}^{(U) }\lambda ^{8}\frac{v}{%
v_\rho},  \notag \\
\varepsilon _{ij}^{(D) } &=&b_{ij}^{(D) }\lambda ^{7}\frac{v}{v_\rho},%
\hspace{0.5cm}i,j=1,2,3;\, n,m=1,2,  \label{bechnmarkquark}
\end{eqnarray}
where $v=\sqrt{v_\rho^2 +v_\eta ^2 }=246$ GeV is the electroweak symmetry
breaking scale. We use the Wolfenstein parameter $\lambda =0.225$ for
characterization of the hierarchy between the parameters defining mass
matrix elements. As discussed above, we consider the hierarchy, which stems
from the model structure rather than from strong tuning of the dimensionless
couplings. Then the coefficients $b_{nm}^{(U) }$ and $b_{ij}^{(D) }$,
constructed from the Yukawa and quartic couplings, are $\mathcal{O}(1)$%
-numbers. In the scenario (\ref{bechnmarkquark}) the exotic quarks $\tilde{T}
$ and $B$ are heavier than the scale of the first stage of the symmetry
breaking (\ref{SB}). As we mentioned earlier, these exotic quarks must be
very heavy for the Universal Seesaw mechanism to operate in our model.

Thus, in the benchmark scenario (\ref{bechnmarkquark}) the SM quark mass
matrices take the form: 
\begin{equation}
\widetilde{M}_U =\left( 
\begin{array}{ccc}
b_{11}^{(U) }\lambda ^{8} & b_{12}^{(U) }\lambda ^{8} & 0 \\ 
b_{21}^{(U) }\lambda ^{7}+a_{21}^{(U) }\lambda ^{4} & b_{22}^{(U) }\lambda
^{7}+a_{22}^{(U) }\lambda ^{4} & 0 \\ 
0 & 0 & m_{t}%
\end{array}%
\right) \frac{v}{\sqrt{2}}=\left( 
\begin{array}{ccc}
b_{11}^{(U) }\lambda^{8} & b_{12}^{(U) }\lambda ^{8} & 0 \\ 
c_{21}^{(U) }\lambda^{4} & c_{22}^{(U) }\lambda ^{4} & 0 \\ 
0 & 0 & m_{t}%
\end{array}%
\right) \frac{v}{\sqrt{2}},
\end{equation}
\begin{equation}
\widetilde{M}_D =\left( 
\begin{array}{ccc}
b_{11}^{(D) }\lambda ^{7} & b_{12}^{(D) }\lambda ^{7} & b_{13}^{(D) }\lambda
^{7} \\ 
b_{21}^{(D) }\lambda ^{7}+a_{21}^{(D) }\lambda ^5 & b_{22}^{(D) }\lambda
^{7}+a_{22}^{(D) }\lambda ^5 & b_{23}^{(D) }\lambda ^{7}+a_{23}^{(D)
}\lambda ^5 \\ 
b_{31}^{(D) }\lambda ^{7}+a_{31}^{(D) }\lambda ^3 & b_{32}^{(D) }\lambda
^{7}+a_{32}^{(D) }\lambda ^3 & b_{33}^{(D) }\lambda ^{7}+a_{33}^{(D)
}\lambda ^3%
\end{array}%
\right) \frac{v}{\sqrt{2}}=\left( 
\begin{array}{ccc}
b_{11}^{(D) }\lambda^7 & b_{12}^{(D) }\lambda ^{7} & b_{13}^{(D) }\lambda^{7}
\\ 
c_{21}^{(D) }\lambda^5 & c_{22}^{(D) }\lambda ^5 & c_{23}^{(D) }\lambda^5 \\ 
c_{31}^{(D) }\lambda^3 & c_{32}^{(D) }\lambda ^3 & c_{33}^{(D) }\lambda^3%
\end{array}%
\right) \frac{v}{\sqrt{2}},
\end{equation}%
where 
\begin{eqnarray}
a_{21}^{(U) } &=&x_{T}z_1 ^{(U) },\hspace{1cm}a_{22}^{(U) }=x_{T}z_2 ^{(U) },
\notag \\
a_{2i}^{(D) } &=&x_B z_{i}^{(D) },\hspace{1cm}a_{3i}^{(D) }=y_B x_{i}^{(D) },%
\hspace{1cm}i=1,2,3.
\end{eqnarray}

The model has 13 dimensionless parameters in the quark sector. This allows
us to reproduce precisely the central experimental values of 10 quark
observables, shown in Table \ref{Observables}. The corresponding values of
the model parameters are:

\begin{eqnarray}
b_{11}^{(U) } &=&c_{21}^{(U) }=1,\hspace{1.3cm}b_{12}^{(U) }\simeq 2.773,%
\hspace{1.5cm}b_{22}^{(U) }\simeq 1.001,\hspace{1.2cm}a_3 ^{(U) }\simeq
0.989,  \notag \\
b_{11}^{(D) } &\simeq &-1.335+0.929i,\hspace{0.1cm}b_{12}^{(D) }\simeq
1.217+1.314 i,\hspace{0.1cm}b_{13}^{(D) }\simeq 2.112-0.929i,  \notag \\
c_{21}^{(D) } &\simeq &-0.869,\hspace{1.4cm}c_{22}^{(D) }\simeq -0.438,%
\hspace{1.2cm}c_{13}^{(D) }\simeq 0.860,  \notag \\
c_{31}^{(D) } &\simeq &-0.707,\hspace{1.4cm}c_{32}^{(D) }\simeq -1.001,%
\hspace{1.2cm}c_{33}^{(D) }\simeq 0.707.
\end{eqnarray}
An important point for us is that all these values are of the order of one.
As we previously discussed, this means that the hierarchy of the quark
masses and mixings originate in our model from its internal structure --
symmetries and field content -- without the need for strong tuning the
dimensionless couplings. 
\begin{table}[tbh]
\begin{center}
\begin{tabular}{c|l}
\hline
Observable & Experimental value \\ \hline
$m_u (MeV)$ & \quad $1.45_{-0.45}^{+0.56}$ \\ \hline
$m_c(MeV)$ & \quad $635\pm 86$ \\ \hline
$m_t(GeV)$ & \quad $172.1\pm 0.6\pm 0.9$ \\ \hline
$m_d(MeV)$ & \quad $2.9_{-0.4}^{+0.5}$ \\ \hline
$m_{s}(MeV)$ & \quad $57.7_{-15.7}^{+16.8}$ \\ \hline
$m_{b}(GeV)$ & \quad $2.82_{-0.04}^{+0.09}$ \\ \hline
$\sin \theta _{12}^{(q)}$ & \quad $0.225$ \\ \hline
$\sin \theta _{23}^{(q)}$ & \quad $0.0421$ \\ \hline
$\sin \theta _{13}^{(q)}$ & \quad $0.00365$ \\ \hline
$J$ & \quad $\left( 3.18\pm 0.15\right) \times 10^{-5}$ \\ \hline
\end{tabular}%
\end{center}
\caption{Experimental $M_{Z}$-scale values of the quark masses \protect\cite%
{Bora:2012tx,Xing:2007fb} and CKM parameters \protect\cite{Tanabashi:2018oca}%
.}
\label{Observables}
\end{table}

\section{Charged Lepton masses and mixings}

\begin{figure}[h]
\resizebox{16cm}{22cm}{\vspace{-8cm}%
\includegraphics{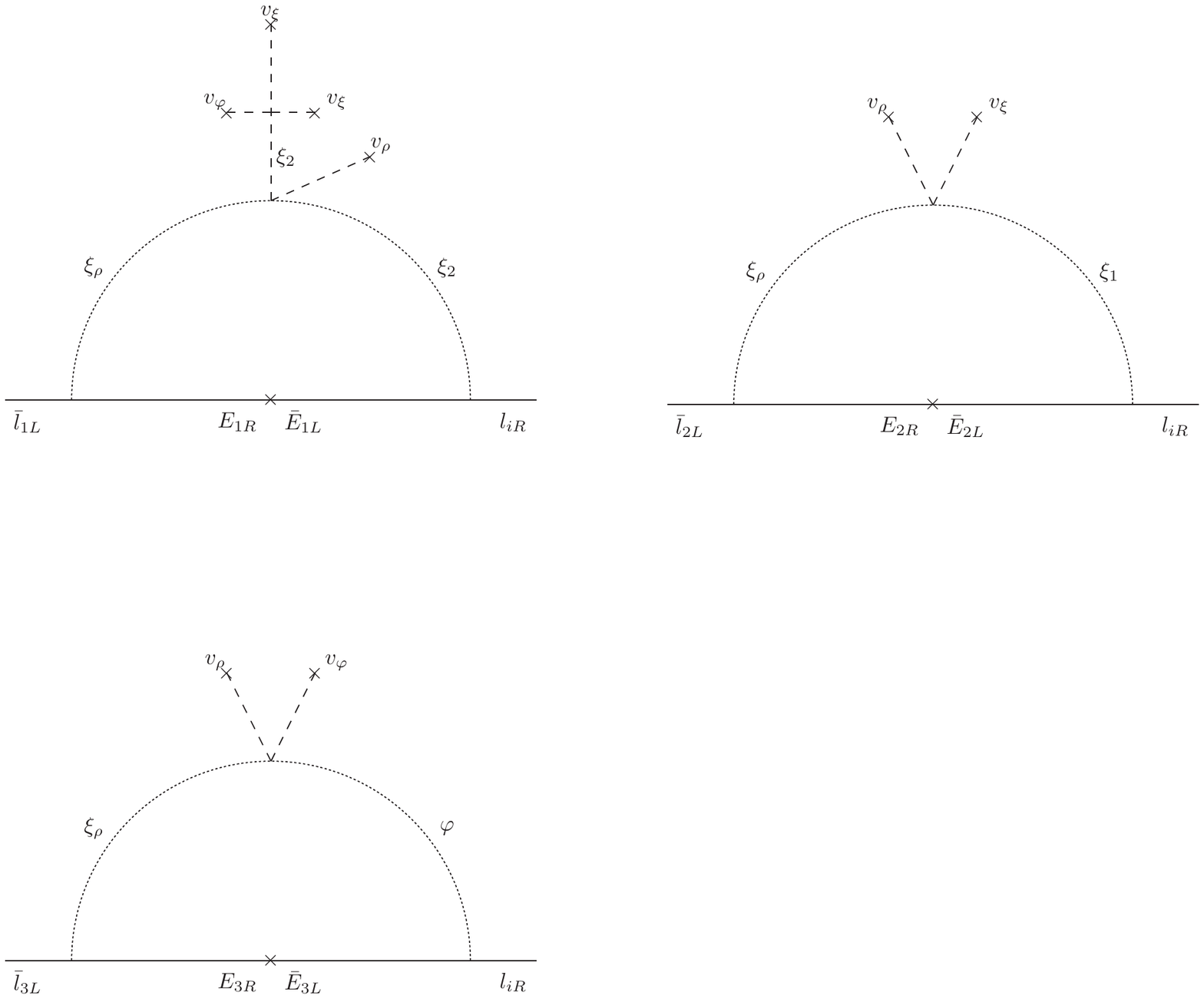}}\vspace{-6cm}
\caption{One-loop Feynman diagrams contributing to the entries of the SM
charged lepton mass matrix. Here $i=1,2,3$.}
\label{Loopdiagramsl}
\end{figure}

%\section{Lepton masses and mixings}
\label{leptonsector} From the charged lepton Yukawa interactions in Eq. (\ref%
{lyl}) we find the charged lepton mass matrix $M_{l}$ in the basis $(%
\overline{l}_{1L}, \overline{l}_{2L},\overline{l}_{3L},\overline{E}_{1L},%
\overline{E}_{2L}, \overline{E}_{3L})$ versus $%
(l_{1R},l_{2R},l_{3R},E_{1R},E_{2R},E_{3R})$ given by: 
\begin{eqnarray}
M_l &=&\left( 
\begin{array}{cc}
\Delta _l & A_l \\ 
B_l & \widetilde{M}_E%
\end{array}%
\right),\hspace{0.2cm}\Delta_l=\left( 
\begin{array}{ccc}
\varepsilon_{11}^{(l) } & \varepsilon_{12}^{(l) } & \varepsilon_{13}^{(l) }
\\ 
\varepsilon_{21}^{(l) } & \varepsilon_{22}^{(l) } & \varepsilon_{23}^{(l) }
\\ 
\varepsilon_{31}^{(l) } & \varepsilon _{32}^{(l) } & \varepsilon _{33}^{(l) }%
\end{array}%
\right) \frac{v_\rho}{\sqrt{2}},\hspace{0.2cm} A_l=\left( 
\begin{array}{ccc}
x_E & 0 & 0 \\ 
0 & x_E & 0 \\ 
0 & 0 & y_E%
\end{array}%
\right) \frac{v_\rho}{\sqrt{2}},  \notag \\
B_l&=&\left( 
\begin{array}{ccc}
0 & 0 & 0 \\ 
-z_1 ^{(l) }v_\xi & -z_2 ^{(l) }v_\xi & -z_3 ^{(l) }v_\xi \\ 
x_1 ^{(l) }v_{\varphi } & x_2 ^{(l) }v_{\varphi } & x_3 ^{(l) }v_{\varphi }%
\end{array}%
\right),\hspace{0.5cm}\widetilde{M}_E =\left( 
\begin{array}{ccc}
m_E & 0 & 0 \\ 
0 & m_E & 0 \\ 
0 & 0 & m_{E_3 }%
\end{array}%
\right),  \notag \\
\varepsilon _{1i}^{(l) } &=&\frac{1}{16\pi ^2 }\frac{\lambda _{\rho ^\dagger
\rho \xi ^2 }\lambda _{\xi ^3 \varphi }x_E z_{i}^{(l) }v_{\varphi }v_\xi ^2 
}{m_E m_{\xi _2 }^2 }\left[ C_{0}\left( \frac{m_{\xi _\rho}}{m_E },\frac{m_{%
\func{Re}\xi _2 }}{m_E }\right) -C_{0}\left( \frac{m_{\zeta _\rho}}{m_E },%
\frac{m_{\func{Im}\xi _2 }}{m_E }\right) \right] ,  \notag \\
\varepsilon _{2i}^{(l) } &=&\frac{1}{16\pi ^2 }\frac{\lambda _{\rho ^\dagger
\rho \xi ^2 }x_E z_{i}^{(l) }v_\xi }{m_E }\left[ C_{0}\left( \frac{m_{\xi
_\rho}}{m_E },\frac{m_{\func{Re}\xi _2 }}{m_E }\right) -C_{0}\left( \frac{%
m_{\zeta _\rho}}{m_E },\frac{m_{\func{Im}\xi _2 }}{m_E }\right) \right] , 
\notag \\
\varepsilon _{3i}^{(l) } &=&\frac{1}{16\pi ^2 }\frac{\lambda _{\rho ^\dagger
\rho \varphi ^2 }x_E z_{i}^{\left( E\right) }v_{\varphi }}{m_E }\left[
C_{0}\left( \frac{m_{\xi _\rho}}{m_E },\frac{m_{\func{Re}\varphi }}{m_E }%
\right) -C_{0}\left( \frac{m_{\zeta _\rho}}{m_E },\frac{m_{\func{Im}\varphi }%
}{m_E }\right) \right],\hspace{0.2cm}i=1,2,3.  \label{Ml}
\end{eqnarray}
where as seen from Eq. (\ref{Ml}), the $\Delta _l$ submatrix is generated at
one loop level %:{\color{blue}
according to the Feynman diagrams shown in Fig.~\ref{Loopdiagramsl} %}.

As follows from Eqs. (\ref{Ml}), the very heavy vector like charged leptons
mix with the SM charged leptons. %{\color{blue}
The former %Such very heavy vector like charged leptons
are assumed to have masses much larger than the $SU(3) _L\times U(1) _X$
symmetry breaking scale. Therefore, analogously to the quark sector, the tau
and muon masses are generated by the tree level Universal seesaw mechanism,
while the electron mass arises from the one loop level radiative seesaw
mechanism. Consequently, 
%we get the following SM charged lepton mass matrix:
SM charged lepton mass matrix takes the form %}
\begin{equation}
\widetilde{M}_l=\Delta _l+A_l\widetilde{M}_E ^{-1}B_l=\left( 
\begin{array}{ccc}
\varepsilon _{11}^{(l) }\frac{v_\rho}{\sqrt{2}} & \varepsilon _{12}^{(l) }%
\frac{v_\rho}{\sqrt{2}} & \varepsilon _{13}^{(l) }\frac{v_\rho}{\sqrt{2}} \\ 
\varepsilon _{21}^{(l) }\frac{v_\rho}{\sqrt{2}}-x_E z_1 ^{(l) }\frac{v_\xi
v_\rho}{\sqrt{2}m_E } & \varepsilon _{22}^{(l) }\frac{v_\rho}{\sqrt{2}}-x_E
z_2 ^{(l) }\frac{v_\xi v_\rho}{\sqrt{2}m_E } & \varepsilon _{23}^{(l) }\frac{%
v_\rho}{\sqrt{2}}-x_E z_3 ^{(l) }\frac{v_\xi v_\rho}{\sqrt{2}m_E } \\ 
\varepsilon _{31}^{(l) }\frac{v_\rho}{\sqrt{2}}+y_E x_1 ^{(l) }\frac{%
v_{\varphi }v_\rho}{\sqrt{2}m_{E_3 }} & \varepsilon _{32}^{(l) }\frac{v_\rho%
}{\sqrt{2}}+y_E x_2 ^{(l) }\frac{v_{\varphi }v_\rho}{\sqrt{2}m_{E_3 }} & 
\varepsilon _{33}^{(l) }\frac{v_\rho}{\sqrt{2}}+y_E x_3 ^{(l) }\frac{%
v_{\varphi }v_\rho}{\sqrt{2}m_{E_3 }}%
\end{array}
\right) .  \label{eq:Charge-SM-leptons}
\end{equation}
In order to show that our model can naturally accommodate the experimental
values of the charged lepton masses we use the extended benchmark scenario (%
\ref{bechnmarkquark}) assuming $m_{E} = m_{\tilde{T}}, m_{E_{3}}=m_{B_{3}}$.
Then we have 
\begin{eqnarray}
v_\xi &=&\lambda ^5\frac{vm_E }{v_\rho},\hspace{1cm}v_{\varphi }=\lambda ^3 
\frac{vm_{E_3 }}{v_\rho},\hspace{1cm}\varepsilon _{ij}^{(l) }=b_{ij}^{(l)
}\lambda ^9\frac{v}{v_\rho},\hspace{1cm}i,j=1,2,3.  \label{bechnmarklepton}
\end{eqnarray}
Here the one-loop contributions $\varepsilon^{(l)}$ are estimated from their
definitions in (\ref{Ml}). Accordingly, the coefficients $b^{(l)}$ are
constructed from the Yukawa and scalar quartic coupling. Thus, in the
benchmark scenario (\ref{bechnmarklepton}) the SM charged lepton mass matrix
reads: 
\begin{equation}
\widetilde{M}_l=\left( 
\begin{array}{ccc}
b_{11}^{(l) }\lambda ^9 & b_{12}^{(l) }\lambda ^9 & b_{13}^{(l) }\lambda ^9
\\ 
b_{21}^{(l) }\lambda ^9+a_{21}^{(l) }\lambda ^5 & b_{22}^{(l) }\lambda
^9+a_{22}^{(l) }\lambda ^5 & b_{23}^{(l) }\lambda ^9+a_{23}^{(l) }\lambda ^5
\\ 
b_{31}^{(l) }\lambda ^9+a_{31}^{(l) }\lambda ^3 & b_{32}^{(l) }\lambda
^9+a_{32}^{(l) }\lambda ^3 & b_{33}^{(l) }\lambda ^9+a_{33}^{(l) }\lambda ^3%
\end{array}%
\right) \frac{v}{\sqrt{2}}=\left( 
\begin{array}{ccc}
c_{11}^{(l) }\lambda ^9 & c_{12}^{(l) }\lambda ^9 & c_{13}^{(l) }\lambda ^9
\\ 
c_{21}^{(l) }\lambda ^5 & c_{22}^{(l) }\lambda ^5 & c_{23}^{(l) }\lambda ^5
\\ 
c_{31}^{(l) }\lambda ^3 & c_{32}^{(l) }\lambda ^3 & c_{33}^{(l) }\lambda ^3%
\end{array}%
\right) \frac{v}{\sqrt{2}},  \label{Mclep}
\end{equation}
where 
\begin{eqnarray}
a^{(l)}_{21}&=&-x_E z_1 ^{(l)},\hspace{1cm} a^{(l)}_{22}=-x_E z_2 ^{(l) }, 
\hspace{1cm} a^{(l)}_{23}=-x_E z_3 ^{(l)}, \\
a^{(l)}_{31}&=&y_E x_1 ^{(l) },\hspace{1cm}\,\,\,\, a^{(l)}_{32}=y_E x_2
^{(l) },\hspace{1cm}\,\,\,\,\,\, a_{33}=y_E x_3 ^{\left( l\right) }, \\
c_{21}^{(l) } &=&b_{21}^{(l) }\lambda^{4}+a_{21}^{(l) },\ \ c_{22}^{(l)
}=b_{22}^{(l) }\lambda ^{4}+a_{22}^{(l) }, \ \ c_{23}^{(l) }=b_{23}^{(l)
}\lambda ^{4}+a_{23}^{(l) }, \\
c_{31}^{(l) } &=&b_{31}^{(l) }\lambda ^{6}+a_{31}^{(l) },\ \ c_{32}^{(l)
}=b_{32}^{(l) }\lambda ^{6}+a_{32}^{(l) },\ \ c_{33}^{(l) }=b_{33}^{(l)
}\lambda ^{6}+a_{33}^{(l) },\hspace{1cm}c^{(l) }_{1i}=b^{(l) }_{1i},\hspace{%
0.5cm}i=1,2,3.  \label{bij}
\end{eqnarray}
The matrix in the second equality of Eq. (\ref{Mclep}) is shown for
convenience in order to explicitly display the hierarchy of the matrix
elements of $\widetilde{M}_l$. To fit the measured values of the SM charged
lepton masses \cite{Tanabashi:2018oca}, we solve the eigenvalue problem for
the SM lepton mass matrix (\ref{Mclep}) and find the following solution: 
\begin{equation}
c_{ij}^{(l)}=\left( 
\begin{array}{ccc}
-1.13637 & -1.03665 & -0.866907 \\ 
-0.658689 & -0.525883 & 1.08155 \\ 
0.900883 & -0.32514 & -0.312796 \\ 
&  & 
\end{array}
\right)
\end{equation}
An important point is that all the elements of this matrix constructed from
Yukawa couplings are $\sim O (1) $. This means that the observed
hierarchical charged lepton mass spectrum can be naturally reproduced in our
model without significant tuning of the coupling constants. 

\section{Neutrino mass generation}

\label{sec:Neutrino masses and mixings} The neutrino Yukawa interactions
give rise to the following neutrino mass terms:

\begin{equation}
-\mathcal{L}_{\text{mass}}^{(\nu )}=\dfrac{1}{2}%
\begin{pmatrix}
\overline{\nu _L^C} & \overline{\nu _R } & \overline{N_R }%
\end{pmatrix}%
M_{\nu }%
\begin{pmatrix}
\nu _L \\ 
\nu _R ^C \\ 
N_R ^C%
\end{pmatrix}%
+H.c,  \label{Lnumass}
\end{equation}

where the neutrino mass matrix $M_{\nu }$ is

\begin{equation}
M_{\nu }=%
\begin{pmatrix}
M_{1} & 0_{3\times 3} & 0_{3\times 3} \\ 
0_{3\times 3} & M_{2} & M_{\chi } \\ 
0_{3\times 3} & M_{\chi }^{T} & \mu%
\end{pmatrix}%
,  \label{Mnu}
\end{equation}%
with the submatrices $M_{1}$ and $M_{2}$ generated at one loop level,
whereas the submatrices $M_{\chi }$ and $\mu $ appearing at tree level. They
are given by:%
\begin{eqnarray}
M_{\chi } &=&\left( 
\begin{array}{ccc}
x_{N} & 0 & 0 \\ 
0 & x_{N} & 0 \\ 
0 & 0 & y_{N}%
\end{array}%
\right) \frac{v_{\chi }}{\sqrt{2}},  \notag \\
\mu &=&\left( 
\begin{array}{ccc}
h_{3N}v_{\sigma _{3}}-h_{1N}v_{\sigma _{1}} & h_{1N}v_{\sigma _{2}} & 
h_{2N}v_{\sigma _{2}} \\ 
h_{1N}v_{\sigma _{2}} & h_{3N}v_{\sigma _{3}}+h_{1N}v_{\sigma _{1}} & 
-h_{2N}v_{\sigma _{1}} \\ 
h_{2N}v_{\sigma _{2}} & -h_{2N}v_{\sigma _{1}} & h_{4N}v_{\sigma _{3}}%
\end{array}%
\right) .  \label{Mnusubmatrices}
\end{eqnarray}%
\newline

\begin{figure}[h]
\includegraphics[width=1.2\textwidth]{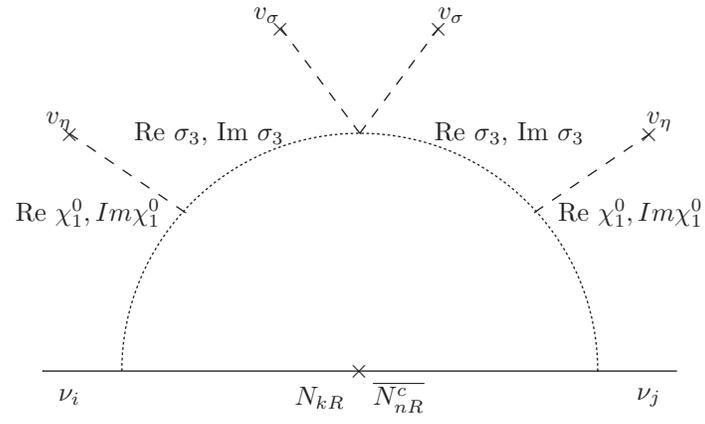} \vspace{-14cm}
\caption{One-loop Feynman diagrams contributing to the entries of the light
active neutrino mass matrix. Here $i,j,k,n=1,2,3$.}
\label{Loopdiagramsmu}
\end{figure}

The light active neutrino mass matrix is generated by the loop diagrams
shown in Figure \ref{Loopdiagramsmu} and is given by: 
\begin{equation}
\widetilde{M}_{\nu }=M_{1}=\left( 
\begin{array}{ccc}
x_{N}^{2}F\left( \mu _{22},m_{\chi _{1R}},m_{\chi _{1I}}\right) \mu _{22} & 
-x_{N}^{2}F\left( \mu _{12},m_{\chi _{1R}},m_{\chi _{1I}}\right) \mu _{12} & 
x_{N}y_{N}F\left( \mu _{23},m_{\chi _{1R}},m_{\chi _{1I}}\right) \mu _{23}
\\ 
-x_{N}^{2}F\left( \mu _{12},m_{\chi _{1R}},m_{\chi _{1I}}\right) \mu _{12} & 
x_{N}^{2}F\left( \mu _{11},m_{\chi _{1R}},m_{\chi _{1I}}\right) \mu _{11} & 
-x_{N}y_{N}F\left( \mu _{13},m_{\chi _{1R}},m_{\chi _{1I}}\right) \mu _{13}
\\ 
x_{N}y_{N}F\left( \mu _{23},m_{\chi _{1R}},m_{\chi _{1I}}\right) \mu _{23} & 
-x_{N}y_{N}F\left( \mu _{13},m_{\chi _{1R}},m_{\chi _{1I}}\right) \mu _{13}
& y_{N}^{2}F\left( \mu _{33},m_{\chi _{1R}},m_{\chi _{1I}}\right) \mu _{33}%
\end{array}%
\right) ,  \label{Mnu}
\end{equation}%
with the loop function of the form \cite{Ma:2006km}: 
\begin{equation}
F\left( m_{1},m_{2},m_{3}\right) =\frac{1}{16\pi ^{2}}\left[ \frac{m_{2}^{2}%
}{m_{2}^{2}-m_{1}^{2}}\ln \left( \frac{m_{2}^{2}}{m_{1}^{2}}\right) -\frac{%
m_{3}^{2}}{m_{3}^{2}-m_{1}^{2}}\ln \left( \frac{m_{3}^{2}}{m_{1}^{2}}\right) %
\right] {\small .}  \label{F}
\end{equation}%
In the limit where $\mu _{ij}^{2}\ll m_{\chi _{1R}}^{2}$, $m_{\chi
_{1I}}^{2} $, the light active neutrino mass matrix becomes: 
\begin{equation}
\widetilde{M}_{\nu }\simeq \frac{m_{\chi _{1R}}^{2}-m_{\chi _{1I}}^{2}}{8\pi
^{2}\left( m_{\chi _{1R}}^{2}+m_{\chi _{1I}}^{2}\right) }\left( 
\begin{array}{ccc}
x_{N}^{2}\mu _{22} & x_{N}^{2}\mu _{12} & x_{N}y_{N}\mu _{23} \\ 
x_{N}^{2}\mu _{12} & x_{N}^{2}\mu _{11} & x_{N}y_{N}\mu _{13} \\ 
x_{N}y_{N}\mu _{23} & x_{N}y_{N}\mu _{13} & y_{N}^{2}\mu _{33}%
\end{array}%
\right) .  \label{lightnumass}
\end{equation}%
All the elements of this mass matrix are free parameters and, therefore, our
model does not predict specific values of neutrino masses and mixing.
However, in our model, the small value of the overall neutrino mass scale is
natural. As seen from Eq.~(\ref{lightnumass}), the smallness of this scale
is attributed to a small splitting $\Delta m_{\chi }^{2}$ between the masses
of the $\chi _{1R}$ and $\chi _{1I}$ scalar fields, which originates from
the quartic term $\gamma \left( \chi ^{\dagger }\eta \sigma _{3}\varphi
+h.c\right) $, so that $\Delta m_{\chi }^{2}\sim \gamma v_{\varphi }^{2}$.
(see Appendix \ref{potential}). Requiring smallness of the parameter $\kappa 
$, we must guarantee its stability with respect radiative corrections, i.e.
its technical naturalness. Checking the model Lagrangian, we observe that in
the limit $\gamma \rightarrow 0$ it acquires an extra symmetry, protecting
this parameter from large radiative corrections. Here we do not need to
specify this group completely and just give its minimal non-trivial
subgroup. This is $Z_{3}$ with the field assignment, where all leptonic
fields as well as the scalar fields $\sigma $ and $\sigma _{3}$ have a
charge equal to $\omega =e^{\frac{2\pi i}{3}}$, whereas the remaining fields
are neutral under this symmetry. This symmetry is broken by the coupling $%
\gamma $. Therefore, in our model small masses of the light neutrinos are
technically natural, being protected by this accidental symmetry. As a
result, the components $\chi _{1,2}$ of the scalar $SU(3)_{L}$-triplet can
be sufficiently light to provide a non-trivial phenomenology.

\section{Conclusions}

\label{conclusions} We have constructed a renormalizable theory based on the 
$SU(3)_C\times SU(3)_L\times U(1)_X$ gauge symmetry, supplemented with the
spontaneously broken $U(1)_{L_g}$ global lepton number symmetry and the $S_3
\times Z_2 $ discrete group, consistent with the low energy SM fermion
flavor data. In our model, the particle spectrum of the 3-3-1 model with
right handed Majorana neutrinos is enlarged by the inclusion of gauge
singlet scalars and charged exotic vector like fermions, which are crucial
for the implementation of the tree level Universal seesaw mechanism that
produces the masses for the bottom, strange and charm quarks as well as the
tau and muon lepton masses. The top and exotic quarks obtain their tree
level masses from renormalizable Yukawa interactions, whereas the first
generation SM charged fermion masses are generated from a one loop level
radiative seesaw mechanism. The masses for the light active neutrinos arise
from a radiative seesaw mechanism at one loop level. The natural smallness
of the overall neutrino mass scale is guarantied by an accidental softly
broken symmetry. Our model successfully explains the hierarchy of the
fermion masses and mixings as well as accommodates the current experimental
deviations of the electron and muon anomalous magnetic moments from their SM
values.

\section*{Acknowledgments}

This research has received funding from Fondecyt (Chile), Grants
No.~1170803, No.~1190845, ANID PIA/APOYO AFB180002, %and the Programa de
%Incentivos a la Iniciaci\'on Cient\' {i}fica (PIIC) from USM
the Vietnam National Foundation for Science and Technology Development
(NAFOSTED) under grant number 103.01-2017.341. A.E.C.H is very grateful to
the Institute of Physics, Vietnam Academy of Science and Technology for the
warm hospitality and for financing his visit.

\appendix

\section{The $S_3 $ discrete group}

\label{appen:s3} The $S_3 $ discrete group contains 3 irreducible
representations: $\mathbf{1}$, $\mathbf{1}^{\prime }$ and $\mathbf{2}$.
Considering $\left( x_1 ,x_2 \right) ^T$\ and $\left( y_1 ,y_2 \right) ^T$
as the basis vectors for two $S_3 $ doublets and $(y{\acute{}})$ is an $S_3 $
non trivial singlet, the multiplication rules of the $S_3 $ group for the
case of real representations take the form \cite{Ishimori:2010au}: 
\begin{eqnarray}
&&\left( 
\begin{array}{c}
x_1 \\ 
x_2%
\end{array}%
\right) _{\mathbf{2 }}\otimes \left( 
\begin{array}{c}
y_1 \\ 
y_{2}%
\end{array}%
\right) _{\mathbf{2}}=\left( x_1 y_1 +x_2 y_2 \right) _{\mathbf{1}}+\left(
x_1 y_2 -x_2 y_1 \right) _{\mathbf{1}^{\prime }}+\left( 
\begin{array}{c}
x_2 y_2 -x_1 y_1 \\ 
x_1 y_2 +x_2 y_1%
\end{array}%
\right) _{\mathbf{2}},  \label{6} \\
&&\left( 
\begin{array}{c}
x_1 \\ 
x_2%
\end{array}%
\right) _{\mathbf{2}}\otimes \left( y%
%TCIMACRO{\U{b4}}%
%BeginExpansion
{\acute{}}%
%EndExpansion
\right) _{\mathbf{1}^{\prime }}=\left( 
\begin{array}{c}
-x_2 y%
%TCIMACRO{\U{b4} }%
%BeginExpansion
{\acute{}}
%EndExpansion
\\ 
x_1 y%
%TCIMACRO{\U{b4}}%
%BeginExpansion
{\acute{}}%
%EndExpansion
\end{array}%
\right) _{\mathbf{2}},\hspace{1cm}\left( x%
%TCIMACRO{\U{b4}}%
%BeginExpansion
{\acute{}}%
%EndExpansion
\right) _{\mathbf{1}^{\prime }}\otimes \left( y%
%TCIMACRO{\U{b4}}%
%BeginExpansion
{\acute{}}%
%EndExpansion
\right) _{\mathbf{1}^{\prime }}=\left( x%
%TCIMACRO{\U{b4}}%
%BeginExpansion
{\acute{}}%
%EndExpansion
y%
%TCIMACRO{\U{b4}}%
%BeginExpansion
{\acute{}}%
%EndExpansion
\right) _{\mathbf{1}}.  \label{7}
\end{eqnarray}

%\Antonio{

\section{The scalar potential}

\label{potential} The renormalizable scalar potential of our model takes the
form: 
\begin{eqnarray}
V &=&-\mu _{\chi }^{2}(\chi ^{\dagger }\chi )-\mu _{\eta }^{2}(\eta
^{\dagger }\eta )-\mu _{\rho }^{2}(\rho ^{\dagger }\rho )-\mu _{\xi
}^{2}(\xi \xi )_{\mathbf{1}}-\mu _{\varphi }^{2}\varphi ^{2}-\mu _{\sigma
}^{2}(\sigma ^{\dagger }\sigma )_{\mathbf{1}}-\mu _{\sigma _{3}}^{2}(\sigma
_{3}^{\dagger }\sigma _{3})  \notag \\
&&+\left( \kappa \varphi \eta _{i}\chi _{j}\rho _{k}\varepsilon
^{ijk}+H.c.\right) +\gamma \left( \chi ^{\dagger }\eta \sigma _{3}\varphi
+h.c\right) +\lambda _{1}(\chi ^{\dagger }\chi )(\chi ^{\dagger }\chi
)+\lambda _{2}(\rho ^{\dagger }\rho )(\rho ^{\dagger }\rho )+\lambda
_{3}(\eta ^{\dagger }\eta )(\eta ^{\dagger }\eta )  \notag \\
&&+\lambda _{4}(\chi ^{\dagger }\chi )(\rho ^{\dagger }\rho )+\lambda
_{5}(\chi ^{\dagger }\chi )(\eta ^{\dagger }\eta )+\lambda _{6}(\rho
^{\dagger }\rho )(\eta ^{\dagger }\eta )+\lambda _{7}(\chi ^{\dagger }\eta
)(\eta ^{\dagger }\chi )+\lambda _{8}(\chi ^{\dagger }\rho )(\rho ^{\dagger
}\chi )+\lambda _{9}(\rho ^{\dagger }\eta )(\eta ^{\dagger }\rho )  \notag \\
&&+\lambda _{10}\left[ (\xi \xi )_{\mathbf{1}}\right] ^{2}+\lambda _{11}%
\left[ (\xi \xi )_{\mathbf{1}^{\prime }}\right] ^{2}+\lambda _{12}\left[
(\xi \xi )_{\mathbf{2}}(\xi \xi )_{\mathbf{2}}\right] _{\mathbf{1}}+\lambda
_{13}\varphi ^{4}+\lambda _{14}\left[ (\sigma ^{\dagger }\sigma )_{\mathbf{1}%
}\right] ^{2}+\lambda _{15}\left[ (\sigma ^{\dagger }\sigma )_{\mathbf{1}%
^{\prime }}\right] ^{2}+\lambda _{16}\left[ (\sigma ^{\dagger }\sigma )_{%
\mathbf{2}}\right] ^{2}  \notag \\
&&+\lambda _{17}(\sigma _{3}^{\dagger }\sigma _{3})^{2}+\lambda _{18}(\xi
\xi )_{\mathbf{1}}\varphi ^{2}+\lambda _{19}\left[ (\xi \xi )_{\mathbf{1}}%
\right] (\sigma ^{\dagger }\sigma )_{\mathbf{1}}+\lambda _{20}\left[ (\xi
\xi )_{\mathbf{1}^{\prime }}\right] (\sigma ^{\dagger }\sigma )_{\mathbf{1}%
^{\prime }}+\lambda _{21}\left[ (\xi \xi )_{\mathbf{2}}\right] (\sigma
^{\dagger }\sigma )_{\mathbf{2}}  \notag \\
&&+\lambda _{22}(\xi \xi )_{\mathbf{1}}(\sigma _{3}^{\dagger }\sigma
_{3})+\lambda _{23}\varphi ^{2}(\sigma ^{\dagger }\sigma )_{\mathbf{1}%
}+\lambda _{24}(\sigma ^{\dagger }\sigma )_{\mathbf{1}}(\sigma _{3}^{\dagger
}\sigma _{3})+\lambda _{25}\left[ (\sigma \sigma )_{\mathbf{1}}(\sigma
_{3}^{\dagger }\sigma _{3}^{\dagger })+h.c\right] +\lambda _{26}\varphi
^{2}(\chi ^{\dagger }\chi )  \notag \\
&&+\lambda _{27}(\xi \xi )_{\mathbf{1}}(\chi ^{\dagger }\chi )+\lambda
_{28}(\sigma ^{\dagger }\sigma )_{\mathbf{1}}(\chi ^{\dagger }\chi )+\lambda
_{29}(\sigma _{3}^{\dagger }\sigma _{3})(\chi ^{\dagger }\chi )+\lambda
_{30}(\xi \xi )_{\mathbf{1}}(\eta ^{\dagger }\eta )+\lambda _{31}\varphi
^{2}(\eta ^{\dagger }\eta )  \notag \\
&&+\lambda _{32}(\sigma ^{\dagger }\sigma )_{\mathbf{1}}(\eta ^{\dagger
}\eta )+\lambda _{33}(\sigma _{3}^{\dagger }\sigma _{3})(\eta ^{\dagger
}\eta )+\lambda _{34}(\xi \xi )_{\mathbf{1}}(\rho ^{\dagger }\rho )+\lambda
_{35}\varphi ^{2}(\rho ^{\dagger }\rho )+\lambda _{36}(\sigma ^{\dagger
}\sigma )_{\mathbf{1}}(\rho ^{\dagger }\rho )  \notag \\
&&+\lambda _{37}(\sigma _{3}^{\dagger }\sigma _{3})(\rho ^{\dagger }\rho
)+\lambda _{38}\left\{ \left[ \left( \xi \xi \right) _{\mathbf{2}}\xi \right]
_{\mathbf{1}^{\prime }}\varphi +h.c\right\} +\lambda _{39}\left[ (\xi \xi )_{%
\mathbf{2}}(\sigma ^{\dagger }\sigma _{3})+h.c\right] +\lambda _{40}\left[
(\sigma ^{\dagger }\xi )_{\mathbf{1}^{\prime }}\varphi \sigma _{3}+h.c\right]
\label{V}
\end{eqnarray}%
where $\lambda _{i}$ ($i=1,2,\cdots ,40$) are dimensionless parameters,
whereas $\mu _{\chi }$, $\mu _{\eta }$, $\mu _{\xi }$, $\mu _{\varphi }$, $%
\mu _{\sigma }$, $\mu _{\sigma _{3}}$, $f$\ \ and $A$ are dimensionful
parameters. Here $\chi $, $\rho $ and $\eta $, the $SU(3)_{L}$ scalar
triplets and the remaining fields are $SU(3)_{L}$ scalar singlets. The
scalar fields $\sigma $ and $\sigma _{3}$ are complex, wheras $\xi $ and $%
\varphi $ are real. %}

% \begin{thebibliography}{9}
% \bibitem{} \input{Biblio10thAugust2020.tex}
% \end{thebibliography}

\section{The scalar potential and scalar mass matrix entries for the g-2
contribution}

\label{appotentialg2}

The scalar potential that contributes to $g-2$ anomalies is given by: {%
\begin{eqnarray}
V^{\text{CP}} &=&-\mu _{\rho }^{2}(\rho ^{\dagger }\rho )-\mu _{\xi
}^{2}(\xi \xi )_{\mathbf{1}}-\mu _{\varphi }^{2}\varphi ^{2}+\left( \kappa
\varphi \eta _{i}\chi _{j}\rho _{k}\varepsilon ^{ijk}+H.c.\right)   \notag
\label{eq:Scalar-CP} \\
&&+\lambda _{2}(\rho ^{\dagger }\rho )(\rho ^{\dagger }\rho )+\lambda _{10}
\left[ (\xi \xi )_{\mathbf{1}}\right] ^{2}+\lambda _{11}\left[ (\xi \xi )_{%
\mathbf{1}^{\prime }}\right] ^{2}  \notag \\
&&+\lambda _{12}\left[ (\xi \xi )_{\mathbf{2}}(\xi \xi )_{\mathbf{2}}\right]
_{\mathbf{1}}+\lambda _{13}\varphi ^{4}+\lambda _{18}(\xi \xi )_{\mathbf{1}%
}\varphi ^{2}  \notag \\
&&+\lambda _{34}(\xi \xi )_{\mathbf{1}}(\rho ^{\dagger }\rho )+\lambda
_{35}\varphi ^{2}(\rho ^{\dagger }\rho )+\lambda _{38}\left\{ \left[ \left(
\xi \xi \right) _{\mathbf{2}}\xi \right] _{\mathbf{1}^{\prime }}\varphi
+h.c\right\} 
\end{eqnarray}%
with the complex quartic coupling $\kappa =\kappa _{r}+i\kappa
_{i}$ introducing CP-violation in the scalar potential. We need it in order to
mix $\varphi $ with $\zeta _{\rho }$. The scalar potential minimization
equations allow us to express the $\mu _{\rho }$, $\mu _{\xi }$ and $\mu
_{\varphi }$ parameters as follows: 
\begin{align}
\mu _{\rho }^{2}& =\frac{1}{2}\left( \frac{\text{$\kappa _{r}$}v_{\eta
}v_{\varphi }v_{\chi }}{v_{\rho }}+\lambda _{34}v_{\xi }^{2}+2\lambda
_{2}v_{\rho }^{2}+\lambda _{35}v_{\varphi }^{2}\right) 
\label{eq:minimization-1} \\
\mu _{\xi }^{2}& =\frac{1}{2}\left( 2\left( \lambda _{10}+\lambda
_{12}\right) v_{\xi }^{2}+\lambda _{34}v_{\rho }^{2}+\lambda _{18}v_{\varphi
}^{2}\right) \label{eq:minimization-2}  \\
\mu _{\varphi }^{2}& =\frac{1}{2}\left( \frac{\text{$\kappa _{r}$}v_{\eta
}v_{\rho }v_{\chi }}{v_{\varphi }}+\lambda _{18}v_{\xi }^{2}+\lambda
_{35}v_{\rho }^{2}+2\lambda _{13}v_{\varphi }^{2}\right) 
\label{eq:minimization-3} 
\end{align}%
 The potential (\ref{eq:Scalar-CP}) generates the entries of
the scalar mass squared matrix $\mathbf{M}^{2}$. In the basis $\xi _{\rho }$%
, $\xi _{1R}$, $\xi _{2R}$, $\varphi $, $\zeta _{\rho }$ these entries are
given by: 
\begin{align}
\allowdisplaybreaks m_{11}^{2}& =2\lambda _{2}v_{\rho }^{2}-\frac{\text{$%
\kappa _{r}$}v_{\eta }v_{\varphi }v_{\chi }}{2v_{\rho }} \\
m_{12}^{2}& =m_{21}^{2}=\lambda _{34}v_{\xi }v_{\rho } \\
m_{14}^{2}& =m_{41}^{2}=\frac{1}{2}\text{$\kappa _{r}$}v_{\eta }v_{\chi
}+\lambda _{35}v_{\rho }v_{\varphi } \\
m_{22}^{2}& =2\left( \lambda _{10}+\lambda _{12}\right) v_{\xi }^{2} \\
m_{23}^{2}& =m_{32}^{2}=-2\sqrt{2}\lambda _{38}v_{\xi }v_{\varphi } \\
m_{24}^{2}& =m_{42}^{2}=\lambda _{18}v_{\xi }v_{\varphi } \\
m_{33}^{2}& =2\lambda _{10}v_{\xi }^{2}-3\lambda _{12}v_{\xi }^{2}-2\lambda
_{38}v_{\varphi }v_{\xi } \\
m_{34}^{2}& =m_{43}^{2}=-\sqrt{2}\lambda _{38}v_{\xi }^{2} \\
m_{44}^{2}& =2\lambda _{13}v_{\varphi }^{2}-\frac{\text{$\kappa _{r}$}%
v_{\eta }v_{\rho }v_{\chi }}{2v_{\varphi }} \\
m_{45}^{2}& =m_{54}^{2}=-\frac{1}{2}\text{$\kappa _{i}$}v_{\eta }v_{\chi } \\
m_{55}^{2}& =-\frac{\text{$\kappa _{r}$}v_{\eta }v_{\varphi }v_{\chi }}{%
2v_{\rho }}.
\end{align}%
From these expressions one can see that by the appropriate choice of the
signs of the quartic couplings it easy to guarantee positive definiteness of
the mass squared matrix (\ref{eq:M2-S}) and, therefore, that the extremum
conditions (\ref{eq:minimization-1})-(\ref{eq:minimization-3}) correspond to
the minimum of the potential (\ref{eq:Scalar-CP}). }

\bibliographystyle{utphys}
\bibliography{Biblio27thJan2021}

\providecommand{\href}[2]{#2}\begingroup\raggedright\begin{thebibliography}{100}

\bibitem{Georgi:1978bv}
H.~Georgi and A.~Pais, ``{Generalization of Gim: Horizontal and Vertical Flavor
  Mixing},''
\href{http://dx.doi.org/10.1103/PhysRevD.19.2746}{{\em Phys. Rev.} {\bfseries
  D19} (1979) 2746}.
%%CITATION = PHRVA,D19,2746;%%.

\bibitem{Valle:1983dk}
J.~W.~F. Valle and M.~Singer, ``{Lepton Number Violation With Quasi Dirac
  Neutrinos},''
\href{http://dx.doi.org/10.1103/PhysRevD.28.540}{{\em Phys. Rev.} {\bfseries
  D28} (1983) 540}.
%%CITATION = PHRVA,D28,540;%%.

\bibitem{Pisano:1991ee}
F.~Pisano and V.~Pleitez, ``{An SU(3) x U(1) model for electroweak
  interactions},'' \href{http://dx.doi.org/10.1103/PhysRevD.46.410}{{\em Phys.
  Rev.} {\bfseries D46} (1992) 410--417},
\href{http://arxiv.org/abs/hep-ph/9206242}{{\ttfamily arXiv:hep-ph/9206242
  [hep-ph]}}.
%%CITATION = HEP-PH/9206242;%%.

\bibitem{Foot:1992rh}
R.~Foot, O.~F. Hernandez, F.~Pisano, and V.~Pleitez, ``{Lepton masses in an
  SU(3)-L x U(1)-N gauge model},''
  \href{http://dx.doi.org/10.1103/PhysRevD.47.4158}{{\em Phys. Rev.} {\bfseries
  D47} (1993) 4158--4161},
\href{http://arxiv.org/abs/hep-ph/9207264}{{\ttfamily arXiv:hep-ph/9207264
  [hep-ph]}}.
%%CITATION = HEP-PH/9207264;%%.

\bibitem{Frampton:1992wt}
P.~H. Frampton, ``{Chiral dilepton model and the flavor question},''
\href{http://dx.doi.org/10.1103/PhysRevLett.69.2889}{{\em Phys. Rev. Lett.}
  {\bfseries 69} (1992) 2889--2891}.
%%CITATION = PRLTA,69,2889;%%.

\bibitem{Hoang:1996gi}
H.~N. Long, ``{SU(3)-L x U(1)-N model for right-handed neutrino neutral
  currents},'' \href{http://dx.doi.org/10.1103/PhysRevD.54.4691}{{\em Phys.
  Rev.} {\bfseries D54} (1996) 4691--4693},
\href{http://arxiv.org/abs/hep-ph/9607439}{{\ttfamily arXiv:hep-ph/9607439
  [hep-ph]}}.
%%CITATION = HEP-PH/9607439;%%.

\bibitem{Hoang:1995vq}
H.~N. Long, ``{The 331 model with right handed neutrinos},''
  \href{http://dx.doi.org/10.1103/PhysRevD.53.437}{{\em Phys. Rev.} {\bfseries
  D53} (1996) 437--445},
\href{http://arxiv.org/abs/hep-ph/9504274}{{\ttfamily arXiv:hep-ph/9504274
  [hep-ph]}}.
%%CITATION = HEP-PH/9504274;%%.

\bibitem{Foot:1994ym}
R.~Foot, H.~N. Long, and T.~A. Tran, ``{$SU(3)_L \otimes U(1)_N$ and $SU(4)_L
  \otimes U(1)_N$ gauge models with right-handed neutrinos},''
  \href{http://dx.doi.org/10.1103/PhysRevD.50.R34}{{\em Phys. Rev.} {\bfseries
  D50} no.~1, (1994) R34--R38},
\href{http://arxiv.org/abs/hep-ph/9402243}{{\ttfamily arXiv:hep-ph/9402243
  [hep-ph]}}.
%%CITATION = HEP-PH/9402243;%%.

\bibitem{CarcamoHernandez:2005ka}
A.~E. Carcamo~Hernandez, R.~Martinez, and F.~Ochoa, ``{Z and Z' decays with and
  without FCNC in 331 models},''
  \href{http://dx.doi.org/10.1103/PhysRevD.73.035007}{{\em Phys. Rev.}
  {\bfseries D73} (2006) 035007},
\href{http://arxiv.org/abs/hep-ph/0510421}{{\ttfamily arXiv:hep-ph/0510421
  [hep-ph]}}.
%%CITATION = HEP-PH/0510421;%%.

\bibitem{Dong:2010zu}
P.~V. Dong, H.~N. Long, D.~V. Soa, and V.~V. Vien, ``{The 3-3-1 model with
  $S_4$ flavor symmetry},''
  \href{http://dx.doi.org/10.1140/epjc/s10052-011-1544-2}{{\em Eur. Phys. J.}
  {\bfseries C71} (2011) 1544},
\href{http://arxiv.org/abs/1009.2328}{{\ttfamily arXiv:1009.2328 [hep-ph]}}.
%%CITATION = ARXIV:1009.2328;%%.

\bibitem{Dong:2010gk}
P.~V. Dong, L.~T. Hue, H.~N. Long, and D.~V. Soa, ``{The 3-3-1 model with
  $\mathrm{A}_4$ flavor symmetry},''
  \href{http://dx.doi.org/10.1103/PhysRevD.81.053004}{{\em Phys. Rev.}
  {\bfseries D81} (2010) 053004},
\href{http://arxiv.org/abs/1001.4625}{{\ttfamily arXiv:1001.4625 [hep-ph]}}.
%%CITATION = ARXIV:1001.4625;%%.

\bibitem{Dong:2011vb}
P.~V. Dong, H.~N. Long, C.~H. Nam, and V.~V. Vien, ``{The $S_3$ flavor symmetry
  in 3-3-1 models},'' \href{http://dx.doi.org/10.1103/PhysRevD.85.053001}{{\em
  Phys. Rev.} {\bfseries D85} (2012) 053001},
\href{http://arxiv.org/abs/1111.6360}{{\ttfamily arXiv:1111.6360 [hep-ph]}}.
%%CITATION = ARXIV:1111.6360;%%.

\bibitem{Benavides:2010zw}
R.~H. Benavides, W.~A. Ponce, and Y.~Giraldo, ``{$SU(3)_c\otimes SU(3)_L\otimes
  U(1)_X$ models with four families},''
  \href{http://dx.doi.org/10.1103/PhysRevD.82.013004}{{\em Phys. Rev.}
  {\bfseries D82} (2010) 013004},
\href{http://arxiv.org/abs/1006.3248}{{\ttfamily arXiv:1006.3248 [hep-ph]}}.
%%CITATION = ARXIV:1006.3248;%%.

\bibitem{Dong:2012bf}
P.~V. Dong, H.~N. Long, and H.~T. Hung, ``{Question of Peccei-Quinn symmetry
  and quark masses in the economical 3-3-1 model},''
  \href{http://dx.doi.org/10.1103/PhysRevD.86.033002}{{\em Phys. Rev.}
  {\bfseries D86} (2012) 033002},
\href{http://arxiv.org/abs/1205.5648}{{\ttfamily arXiv:1205.5648 [hep-ph]}}.
%%CITATION = ARXIV:1205.5648;%%.

\bibitem{Huong:2012pg}
D.~T. Huong, L.~T. Hue, M.~C. Rodriguez, and H.~N. Long, ``{Supersymmetric
  reduced minimal 3-3-1 model},''
  \href{http://dx.doi.org/10.1016/j.nuclphysb.2013.01.016}{{\em Nucl. Phys.}
  {\bfseries B870} (2013) 293--322},
\href{http://arxiv.org/abs/1210.6776}{{\ttfamily arXiv:1210.6776 [hep-ph]}}.
%%CITATION = ARXIV:1210.6776;%%.

\bibitem{Giang:2012vs}
P.~T. Giang, L.~T. Hue, D.~T. Huong, and H.~N. Long, ``{Lepton-Flavor Violating
  Decays of Neutral Higgs to Muon and Tauon in Supersymmetric Economical 3-3-1
  Model},'' \href{http://dx.doi.org/10.1016/j.nuclphysb.2012.06.008}{{\em Nucl.
  Phys.} {\bfseries B864} (2012) 85--112},
\href{http://arxiv.org/abs/1204.2902}{{\ttfamily arXiv:1204.2902 [hep-ph]}}.
%%CITATION = ARXIV:1204.2902;%%.

\bibitem{Binh:2013axa}
D.~T. Binh, L.~T. Hue, D.~T. Huong, and H.~N. Long, ``{Higgs revised in
  supersymmetric economical 3-3-1 model with $B / \mu$-type terms},''
  \href{http://dx.doi.org/10.1140/epjc/s10052-014-2851-1}{{\em Eur. Phys. J.}
  {\bfseries C74} no.~5, (2014) 2851},
\href{http://arxiv.org/abs/1308.3085}{{\ttfamily arXiv:1308.3085 [hep-ph]}}.
%%CITATION = ARXIV:1308.3085;%%.

\bibitem{Hernandez:2013mcf}
A.~E. Carcamo~Hernandez, R.~Martinez, and F.~Ochoa, ``{Radiative seesaw-type
  mechanism of quark masses in $SU(3)_C \otimes SU(3)_L \otimes U(1)_X$},''
  \href{http://dx.doi.org/10.1103/PhysRevD.87.075009}{{\em Phys. Rev.}
  {\bfseries D87} no.~7, (2013) 075009},
\href{http://arxiv.org/abs/1302.1757}{{\ttfamily arXiv:1302.1757 [hep-ph]}}.
%%CITATION = ARXIV:1302.1757;%%.

\bibitem{Hernandez:2013hea}
A.~E. Cárcamo~Hernández, R.~Martinez, and F.~Ochoa, ``{Fermion masses and
  mixings in the 3-3-1 model with right-handed neutrinos based on the $S_3$
  flavor symmetry},''
  \href{http://dx.doi.org/10.1140/epjc/s10052-016-4480-3}{{\em Eur. Phys. J.}
  {\bfseries C76} no.~11, (2016) 634},
\href{http://arxiv.org/abs/1309.6567}{{\ttfamily arXiv:1309.6567 [hep-ph]}}.
%%CITATION = ARXIV:1309.6567;%%.

\bibitem{Hernandez:2014vta}
A.~E. Cárcamo~Hernández, R.~Martinez, and J.~Nisperuza, ``{$S_3$ discrete
  group as a source of the quark mass and mixing pattern in $331$ models},''
  \href{http://dx.doi.org/10.1140/epjc/s10052-015-3278-z}{{\em Eur. Phys. J.}
  {\bfseries C75} no.~2, (2015) 72},
\href{http://arxiv.org/abs/1401.0937}{{\ttfamily arXiv:1401.0937 [hep-ph]}}.
%%CITATION = ARXIV:1401.0937;%%.

\bibitem{Hernandez:2014lpa}
A.~E. Cárcamo~Hernández, E.~Cataño~Mur, and R.~Martinez, ``{Lepton masses
  and mixing in $SU(3)_{C}\otimes SU(3)_{L}\otimes U(1)_{X}$ models with a
  $S_3$ flavor symmetry},''
  \href{http://dx.doi.org/10.1103/PhysRevD.90.073001}{{\em Phys. Rev.}
  {\bfseries D90} no.~7, (2014) 073001},
\href{http://arxiv.org/abs/1407.5217}{{\ttfamily arXiv:1407.5217 [hep-ph]}}.
%%CITATION = ARXIV:1407.5217;%%.

\bibitem{Kelso:2014qka}
C.~Kelso, H.~N. Long, R.~Martinez, and F.~S. Queiroz, ``{Connection of
  $g-2_{\mu}$, electroweak, dark matter, and collider constraints on 331
  models},'' \href{http://dx.doi.org/10.1103/PhysRevD.90.113011}{{\em Phys.
  Rev.} {\bfseries D90} no.~11, (2014) 113011},
\href{http://arxiv.org/abs/1408.6203}{{\ttfamily arXiv:1408.6203 [hep-ph]}}.
%%CITATION = ARXIV:1408.6203;%%.

\bibitem{Vien:2014gza}
V.~V. Vien and H.~N. Long, ``{The $T_7$ flavor symmetry in 3-3-1 model with
  neutral leptons},'' \href{http://dx.doi.org/10.1007/JHEP04(2014)133}{{\em
  JHEP} {\bfseries 04} (2014) 133},
\href{http://arxiv.org/abs/1402.1256}{{\ttfamily arXiv:1402.1256 [hep-ph]}}.
%%CITATION = ARXIV:1402.1256;%%.

\bibitem{Phong:2014ofa}
V.~Q. Phong, H.~N. Long, V.~T. Van, and L.~H. Minh, ``{Electroweak phase
  transition in the economical 3-3-1 model},''
  \href{http://dx.doi.org/10.1140/epjc/s10052-015-3550-2}{{\em Eur. Phys. J.}
  {\bfseries C75} no.~7, (2015) 342},
\href{http://arxiv.org/abs/1409.0750}{{\ttfamily arXiv:1409.0750 [hep-ph]}}.
%%CITATION = ARXIV:1409.0750;%%.

\bibitem{Phong:2014yca}
V.~Q. Phong, H.~N. Long, V.~T. Van, and N.~C. Thanh, ``{Electroweak sphalerons
  in the reduced minimal 3-3-1 model},''
  \href{http://dx.doi.org/10.1103/PhysRevD.90.085019}{{\em Phys. Rev.}
  {\bfseries D90} no.~8, (2014) 085019},
\href{http://arxiv.org/abs/1408.5657}{{\ttfamily arXiv:1408.5657 [hep-ph]}}.
%%CITATION = ARXIV:1408.5657;%%.

\bibitem{Boucenna:2014ela}
S.~M. Boucenna, S.~Morisi, and J.~W.~F. Valle, ``{Radiative neutrino mass in
  3-3-1 scheme},'' \href{http://dx.doi.org/10.1103/PhysRevD.90.013005}{{\em
  Phys. Rev.} {\bfseries D90} no.~1, (2014) 013005},
\href{http://arxiv.org/abs/1405.2332}{{\ttfamily arXiv:1405.2332 [hep-ph]}}.
%%CITATION = ARXIV:1405.2332;%%.

\bibitem{DeConto:2015eia}
G.~De~Conto, A.~C.~B. Machado, and V.~Pleitez, ``{Minimal 3-3-1 model with a
  spectator sextet},'' \href{http://dx.doi.org/10.1103/PhysRevD.92.075031}{{\em
  Phys. Rev.} {\bfseries D92} no.~7, (2015) 075031},
\href{http://arxiv.org/abs/1505.01343}{{\ttfamily arXiv:1505.01343 [hep-ph]}}.
%%CITATION = ARXIV:1505.01343;%%.

\bibitem{Boucenna:2015zwa}
S.~M. Boucenna, J.~W.~F. Valle, and A.~Vicente, ``{Predicting charged lepton
  flavor violation from 3-3-1 gauge symmetry},''
  \href{http://dx.doi.org/10.1103/PhysRevD.92.053001}{{\em Phys. Rev.}
  {\bfseries D92} no.~5, (2015) 053001},
\href{http://arxiv.org/abs/1502.07546}{{\ttfamily arXiv:1502.07546 [hep-ph]}}.
%%CITATION = ARXIV:1502.07546;%%.

\bibitem{Boucenna:2015pav}
S.~M. Boucenna, S.~Morisi, and A.~Vicente, ``{The LHC diphoton resonance from
  gauge symmetry},'' \href{http://dx.doi.org/10.1103/PhysRevD.93.115008}{{\em
  Phys. Rev.} {\bfseries D93} no.~11, (2016) 115008},
\href{http://arxiv.org/abs/1512.06878}{{\ttfamily arXiv:1512.06878 [hep-ph]}}.
%%CITATION = ARXIV:1512.06878;%%.

\bibitem{Benavides:2015afa}
R.~H. Benavides, L.~N. Epele, H.~Fanchiotti, C.~G. Canal, and W.~A. Ponce,
  ``{Lepton number violation and neutrino masses in 3-3-1 models},''
  \href{http://dx.doi.org/10.1155/2015/813129}{{\em Adv. High Energy Phys.}
  {\bfseries 2015} (2015) 813129},
\href{http://arxiv.org/abs/1503.01686}{{\ttfamily arXiv:1503.01686 [hep-ph]}}.
%%CITATION = ARXIV:1503.01686;%%.

\bibitem{Hernandez:2015tna}
A.~E. Cárcamo~Hernández and R.~Martinez, ``{A predictive 3-3-1 model with
  $A_4$ flavor symmetry},''
  \href{http://dx.doi.org/10.1016/j.nuclphysb.2016.02.025}{{\em Nucl. Phys.}
  {\bfseries B905} (2016) 337--358},
\href{http://arxiv.org/abs/1501.05937}{{\ttfamily arXiv:1501.05937 [hep-ph]}}.
%%CITATION = ARXIV:1501.05937;%%.

\bibitem{Hernandez:2015cra}
A.~E. Cárcamo~Hernández and R.~Martinez, ``{Fermion mass and mixing pattern
  in a minimal T7 flavor 331 model},''
  \href{http://dx.doi.org/10.1088/0954-3899/43/4/045003}{{\em J. Phys.}
  {\bfseries G43} no.~4, (2016) 045003},
\href{http://arxiv.org/abs/1501.07261}{{\ttfamily arXiv:1501.07261 [hep-ph]}}.
%%CITATION = ARXIV:1501.07261;%%.

\bibitem{Hue:2015fbb}
L.~T. Hue, H.~N. Long, T.~T. Thuc, and T.~Phong~Nguyen, ``{Lepton flavor
  violating decays of Standard-Model-like Higgs in 3-3-1 model with neutral
  lepton},'' \href{http://dx.doi.org/10.1016/j.nuclphysb.2016.03.034}{{\em
  Nucl. Phys.} {\bfseries B907} (2016) 37--76},
\href{http://arxiv.org/abs/1512.03266}{{\ttfamily arXiv:1512.03266 [hep-ph]}}.
%%CITATION = ARXIV:1512.03266;%%.

\bibitem{Hernandez:2015ywg}
A.~E.~C. Hernández and I.~Nišandžić, ``{LHC diphoton resonance at 750 GeV
  as an indication of $SU(3)_L\times U(1)_X$ electroweak symmetry},''
  \href{http://dx.doi.org/10.1140/epjc/s10052-016-4230-6}{{\em Eur. Phys. J.}
  {\bfseries C76} no.~7, (2016) 380},
\href{http://arxiv.org/abs/1512.07165}{{\ttfamily arXiv:1512.07165 [hep-ph]}}.
%%CITATION = ARXIV:1512.07165;%%.

\bibitem{Fonseca:2016tbn}
R.~M. Fonseca and M.~Hirsch, ``{A flipped 331 model},''
  \href{http://dx.doi.org/10.1007/JHEP08(2016)003}{{\em JHEP} {\bfseries 08}
  (2016) 003},
\href{http://arxiv.org/abs/1606.01109}{{\ttfamily arXiv:1606.01109 [hep-ph]}}.
%%CITATION = ARXIV:1606.01109;%%.

\bibitem{Vien:2016tmh}
V.~V. Vien, A.~E. Cárcamo~Hernández, and H.~N. Long, ``{The $\Delta(27)$
  flavor 3-3-1 model with neutral leptons},''
  \href{http://dx.doi.org/10.1016/j.nuclphysb.2016.10.010}{{\em Nucl. Phys.}
  {\bfseries B913} (2016) 792--814},
\href{http://arxiv.org/abs/1601.03300}{{\ttfamily arXiv:1601.03300 [hep-ph]}}.
%%CITATION = ARXIV:1601.03300;%%.

\bibitem{Hernandez:2016eod}
A.~E. Cárcamo~Hernández, H.~N. Long, and V.~V. Vien, ``{A 3-3-1 model with
  right-handed neutrinos based on the $\varDelta \left( 27\right) $ family
  symmetry},'' \href{http://dx.doi.org/10.1140/epjc/s10052-016-4074-0}{{\em
  Eur. Phys. J.} {\bfseries C76} no.~5, (2016) 242},
\href{http://arxiv.org/abs/1601.05062}{{\ttfamily arXiv:1601.05062 [hep-ph]}}.
%%CITATION = ARXIV:1601.05062;%%.

\bibitem{Fonseca:2016xsy}
R.~M. Fonseca and M.~Hirsch, ``{Lepton number violation in 331 models},''
  \href{http://dx.doi.org/10.1103/PhysRevD.94.115003}{{\em Phys. Rev.}
  {\bfseries D94} no.~11, (2016) 115003},
\href{http://arxiv.org/abs/1607.06328}{{\ttfamily arXiv:1607.06328 [hep-ph]}}.
%%CITATION = ARXIV:1607.06328;%%.

\bibitem{Deppisch:2016jzl}
F.~F. Deppisch, C.~Hati, S.~Patra, U.~Sarkar, and J.~W.~F. Valle, ``{331 Models
  and Grand Unification: From Minimal SU(5) to Minimal SU(6)},''
  \href{http://dx.doi.org/10.1016/j.physletb.2016.10.002}{{\em Phys. Lett.}
  {\bfseries B762} (2016) 432--440},
\href{http://arxiv.org/abs/1608.05334}{{\ttfamily arXiv:1608.05334 [hep-ph]}}.
%%CITATION = ARXIV:1608.05334;%%.

\bibitem{Reig:2016ewy}
M.~Reig, J.~W.~F. Valle, and C.~A. Vaquera-Araujo, ``{Realistic
  $\mathrm{SU(3)_c \otimes SU(3)_L \otimes U(1)_X}$ model with a type II Dirac
  neutrino seesaw mechanism},''
  \href{http://dx.doi.org/10.1103/PhysRevD.94.033012}{{\em Phys. Rev.}
  {\bfseries D94} no.~3, (2016) 033012},
\href{http://arxiv.org/abs/1606.08499}{{\ttfamily arXiv:1606.08499 [hep-ph]}}.
%%CITATION = ARXIV:1606.08499;%%.

\bibitem{CarcamoHernandez:2017cwi}
A.~E. Cárcamo~Hernández, S.~Kovalenko, H.~N. Long, and I.~Schmidt, ``{A
  variant of 3-3-1 model for the generation of the SM fermion mass and mixing
  pattern},'' \href{http://dx.doi.org/10.1007/JHEP07(2018)144}{{\em JHEP}
  {\bfseries 07} (2018) 144},
\href{http://arxiv.org/abs/1705.09169}{{\ttfamily arXiv:1705.09169 [hep-ph]}}.
%%CITATION = ARXIV:1705.09169;%%.

\bibitem{CarcamoHernandez:2017kra}
A.~E. Cárcamo~Hernández and H.~N. Long, ``{A highly predictive $A_{4}$
  flavour 3-3-1 model with radiative inverse seesaw mechanism},''
  \href{http://dx.doi.org/10.1088/1361-6471/aaace7}{{\em J. Phys.} {\bfseries
  G45} no.~4, (2018) 045001},
\href{http://arxiv.org/abs/1705.05246}{{\ttfamily arXiv:1705.05246 [hep-ph]}}.
%%CITATION = ARXIV:1705.05246;%%.

\bibitem{Hati:2017aez}
C.~Hati, S.~Patra, M.~Reig, J.~W.~F. Valle, and C.~A. Vaquera-Araujo,
  ``{Towards gauge coupling unification in left-right symmetric
  $\mathrm{SU(3)_c \times SU(3)_L \times SU(3)_R \times U(1)_{X}}$ theories},''
  \href{http://dx.doi.org/10.1103/PhysRevD.96.015004}{{\em Phys. Rev.}
  {\bfseries D96} no.~1, (2017) 015004},
\href{http://arxiv.org/abs/1703.09647}{{\ttfamily arXiv:1703.09647 [hep-ph]}}.
%%CITATION = ARXIV:1703.09647;%%.

\bibitem{Barreto:2017xix}
E.~R. Barreto, A.~G. Dias, J.~Leite, C.~C. Nishi, R.~L.~N. Oliveira, and W.~C.
  Vieira, ``{Hierarchical fermions and detectable $Z^{\prime}$ from effective
  two-Higgs-triplet 3-3-1 model},''
  \href{http://dx.doi.org/10.1103/PhysRevD.97.055047}{{\em Phys. Rev.}
  {\bfseries D97} no.~5, (2018) 055047},
\href{http://arxiv.org/abs/1709.09946}{{\ttfamily arXiv:1709.09946 [hep-ph]}}.
%%CITATION = ARXIV:1709.09946;%%.

\bibitem{CarcamoHernandez:2018iel}
A.~E. Cárcamo~Hernández, H.~N. Long, and V.~V. Vien, ``{The first
  $\Delta(27)$ flavor 3-3-1 model with low scale seesaw mechanism},''
  \href{http://dx.doi.org/10.1140/epjc/s10052-018-6284-0}{{\em Eur. Phys. J.}
  {\bfseries C78} no.~10, (2018) 804},
\href{http://arxiv.org/abs/1803.01636}{{\ttfamily arXiv:1803.01636 [hep-ph]}}.
%%CITATION = ARXIV:1803.01636;%%.

\bibitem{Vien:2018otl}
V.~V. Vien, H.~N. Long, and A.~E. Cárcamo~Hernández, ``{Lepton masses and
  mixings in a $T'$ flavoured 3-3-1 model with type I and II seesaw
  mechanisms},'' \href{http://dx.doi.org/10.1142/S0217732319500056}{{\em Mod.
  Phys. Lett.} {\bfseries A34} no.~01, (2019) 1950005},
\href{http://arxiv.org/abs/1812.07263}{{\ttfamily arXiv:1812.07263 [hep-ph]}}.
%%CITATION = ARXIV:1812.07263;%%.

\bibitem{Dias:2018ddy}
A.~G. Dias, J.~Leite, D.~D. Lopes, and C.~C. Nishi, ``{Fermion Mass Hierarchy
  and Double Seesaw Mechanism in a 3-3-1 Model with an Axion},''
  \href{http://dx.doi.org/10.1103/PhysRevD.98.115017}{{\em Phys. Rev.}
  {\bfseries D98} no.~11, (2018) 115017},
\href{http://arxiv.org/abs/1810.01893}{{\ttfamily arXiv:1810.01893 [hep-ph]}}.
%%CITATION = ARXIV:1810.01893;%%.

\bibitem{Ferreira:2019qpf}
M.~M. Ferreira, T.~B. de~Melo, S.~Kovalenko, P.~R.~D. Pinheiro, and F.~S.
  Queiroz, ``{Lepton Flavor Violation and Collider Searches in a Type I + II
  Seesaw Model},'' \href{http://dx.doi.org/10.1140/epjc/s10052-019-7422-z}{{\em
  Eur. Phys. J.} {\bfseries C79} no.~11, (2019) 955},
\href{http://arxiv.org/abs/1903.07634}{{\ttfamily arXiv:1903.07634 [hep-ph]}}.
%%CITATION = ARXIV:1903.07634;%%.

\bibitem{CarcamoHernandez:2019vih}
A.~E. Cárcamo~Hernández, Y.~Hidalgo~Velásquez, and N.~A. Pérez-Julve, ``{A
  3-3-1 model with low scale seesaw mechanisms},''
  \href{http://dx.doi.org/10.1140/epjc/s10052-019-7325-z}{{\em Eur. Phys. J.}
  {\bfseries C79} no.~10, (2019) 828},
\href{http://arxiv.org/abs/1905.02323}{{\ttfamily arXiv:1905.02323 [hep-ph]}}.
%%CITATION = ARXIV:1905.02323;%%.

\bibitem{CarcamoHernandez:2019iwh}
A.~E. Cárcamo~Hernández, N.~A. Pérez-Julve, and Y.~Hidalgo~Velásquez,
  ``{Fermion masses and mixings and some phenomenological aspects of a 3-3-1
  model with linear seesaw mechanism},''
  \href{http://dx.doi.org/10.1103/PhysRevD.100.095025}{{\em Phys. Rev.}
  {\bfseries D100} no.~9, (2019) 095025},
\href{http://arxiv.org/abs/1907.13083}{{\ttfamily arXiv:1907.13083 [hep-ph]}}.
%%CITATION = ARXIV:1907.13083;%%.

\bibitem{CarcamoHernandez:2019lhv}
A.~E. Cárcamo~Hernández, D.~T. Huong, and H.~N. Long, ``{Minimal model for
  the fermion flavor structure, mass hierarchy, dark matter, leptogenesis, and
  the electron and muon anomalous magnetic moments},''
  \href{http://dx.doi.org/10.1103/PhysRevD.102.055002}{{\em Phys. Rev.}
  {\bfseries D102} no.~5, (2020) 055002},
\href{http://arxiv.org/abs/1910.12877}{{\ttfamily arXiv:1910.12877 [hep-ph]}}.
%%CITATION = ARXIV:1910.12877;%%.

\bibitem{CarcamoHernandez:2020pnh}
A.~E. Cárcamo~Hernández, L.~T. Hue, S.~Kovalenko, and H.~N. Long, ``{An
  extended 3-3-1 model with two scalar triplets and linear seesaw mechanism},''
\href{http://arxiv.org/abs/2001.01748}{{\ttfamily arXiv:2001.01748 [hep-ph]}}.
%%CITATION = ARXIV:2001.01748;%%.

\bibitem{Crivellin:2018qmi}
A.~Crivellin, M.~Hoferichter, and P.~Schmidt-Wellenburg, ``{Combined
  explanations of $(g-2)_{\mu,e}$ and implications for a large muon EDM},''
  \href{http://dx.doi.org/10.1103/PhysRevD.98.113002}{{\em Phys. Rev.}
  {\bfseries D98} no.~11, (2018) 113002},
\href{http://arxiv.org/abs/1807.11484}{{\ttfamily arXiv:1807.11484 [hep-ph]}}.
%%CITATION = ARXIV:1807.11484;%%.

\bibitem{Endo:2019bcj}
M.~Endo and W.~Yin, ``{Explaining electron and muon $g-2$ anomaly in SUSY
  without lepton-flavor mixings},''
  \href{http://dx.doi.org/10.1007/JHEP08(2019)122}{{\em JHEP} {\bfseries 08}
  (2019) 122},
\href{http://arxiv.org/abs/1906.08768}{{\ttfamily arXiv:1906.08768 [hep-ph]}}.
%%CITATION = ARXIV:1906.08768;%%.

\bibitem{Giudice:2012ms}
G.~F. Giudice, P.~Paradisi, and M.~Passera, ``{Testing new physics with the
  electron g-2},'' \href{http://dx.doi.org/10.1007/JHEP11(2012)113}{{\em JHEP}
  {\bfseries 11} (2012) 113},
\href{http://arxiv.org/abs/1208.6583}{{\ttfamily arXiv:1208.6583 [hep-ph]}}.
%%CITATION = ARXIV:1208.6583;%%.

\bibitem{Falkowski:2018dsl}
A.~Falkowski, S.~F. King, E.~Perdomo, and M.~Pierre, ``{Flavourful $Z'$ portal
  for vector-like neutrino Dark Matter and $R_{K^{(*)}}$},''
  \href{http://dx.doi.org/10.1007/JHEP08(2018)061}{{\em JHEP} {\bfseries 08}
  (2018) 061},
\href{http://arxiv.org/abs/1803.04430}{{\ttfamily arXiv:1803.04430 [hep-ph]}}.
%%CITATION = ARXIV:1803.04430;%%.

\bibitem{Allanach:2015gkd}
B.~Allanach, F.~S. Queiroz, A.~Strumia, and S.~Sun, ``{$Z^{\prime}$ models for
  the LHCb and $g-2$ muon anomalies},''
  \href{http://dx.doi.org/10.1103/PhysRevD.93.055045,
  10.1103/PhysRevD.95.119902}{{\em Phys. Rev.} {\bfseries D93} no.~5, (2016)
  055045}, \href{http://arxiv.org/abs/1511.07447}{{\ttfamily arXiv:1511.07447
  [hep-ph]}}.
[Erratum: Phys. Rev.D95,no.11,119902(2017)].
%%CITATION = ARXIV:1511.07447;%%.

\bibitem{Chen:2016dip}
C.-H. Chen, T.~Nomura, and H.~Okada, ``{Explanation of $B \to K^{(*)} \ell^+
  \ell^-$ and muon $g-2$, and implications at the LHC},''
  \href{http://dx.doi.org/10.1103/PhysRevD.94.115005}{{\em Phys. Rev.}
  {\bfseries D94} no.~11, (2016) 115005},
\href{http://arxiv.org/abs/1607.04857}{{\ttfamily arXiv:1607.04857 [hep-ph]}}.
%%CITATION = ARXIV:1607.04857;%%.

\bibitem{Raby:2017igl}
S.~Raby and A.~Trautner, ``{Vectorlike chiral fourth family to explain muon
  anomalies},'' \href{http://dx.doi.org/10.1103/PhysRevD.97.095006}{{\em Phys.
  Rev.} {\bfseries D97} no.~9, (2018) 095006},
\href{http://arxiv.org/abs/1712.09360}{{\ttfamily arXiv:1712.09360 [hep-ph]}}.
%%CITATION = ARXIV:1712.09360;%%.

\bibitem{Chiang:2017tai}
C.-W. Chiang, H.~Okada, and E.~Senaha, ``{Dark matter, muon $g-2$, electric
  dipole moments, and $Z\to \ell_i^+ \ell_j^-$ in a one-loop induced neutrino
  model},'' \href{http://dx.doi.org/10.1103/PhysRevD.96.015002}{{\em Phys.
  Rev.} {\bfseries D96} no.~1, (2017) 015002},
\href{http://arxiv.org/abs/1703.09153}{{\ttfamily arXiv:1703.09153 [hep-ph]}}.
%%CITATION = ARXIV:1703.09153;%%.

\bibitem{Chen:2017hir}
C.-H. Chen, T.~Nomura, and H.~Okada, ``{Excesses of muon $g-2$,
  $R_{D^{(\ast)}}$, and $R_K$ in a leptoquark model},''
  \href{http://dx.doi.org/10.1016/j.physletb.2017.10.005}{{\em Phys. Lett.}
  {\bfseries B774} (2017) 456--464},
\href{http://arxiv.org/abs/1703.03251}{{\ttfamily arXiv:1703.03251 [hep-ph]}}.
%%CITATION = ARXIV:1703.03251;%%.

\bibitem{Davoudiasl:2018fbb}
H.~Davoudiasl and W.~J. Marciano, ``{Tale of two anomalies},''
  \href{http://dx.doi.org/10.1103/PhysRevD.98.075011}{{\em Phys. Rev.}
  {\bfseries D98} no.~7, (2018) 075011},
\href{http://arxiv.org/abs/1806.10252}{{\ttfamily arXiv:1806.10252 [hep-ph]}}.
%%CITATION = ARXIV:1806.10252;%%.

\bibitem{Liu:2018xkx}
J.~Liu, C.~E.~M. Wagner, and X.-P. Wang, ``{A light complex scalar for the
  electron and muon anomalous magnetic moments},''
  \href{http://dx.doi.org/10.1007/JHEP03(2019)008}{{\em JHEP} {\bfseries 03}
  (2019) 008},
\href{http://arxiv.org/abs/1810.11028}{{\ttfamily arXiv:1810.11028 [hep-ph]}}.
%%CITATION = ARXIV:1810.11028;%%.

\bibitem{CarcamoHernandez:2019xkb}
A.~E. Cárcamo~Hernández, S.~Kovalenko, R.~Pasechnik, and I.~Schmidt,
  ``{Phenomenology of an extended IDM with loop-generated fermion mass
  hierarchies},'' \href{http://dx.doi.org/10.1140/epjc/s10052-019-7101-0}{{\em
  Eur. Phys. J.} {\bfseries C79} no.~7, (2019) 610},
\href{http://arxiv.org/abs/1901.09552}{{\ttfamily arXiv:1901.09552 [hep-ph]}}.
%%CITATION = ARXIV:1901.09552;%%.

\bibitem{Nomura:2019btk}
T.~Nomura and H.~Okada, ``{Muon anomalous magnetic moment, $Z$ boson decays,
  and collider physics in multicharged particles},''
  \href{http://dx.doi.org/10.1103/PhysRevD.101.015021}{{\em Phys. Rev.}
  {\bfseries D101} no.~1, (2020) 015021},
\href{http://arxiv.org/abs/1903.05958}{{\ttfamily arXiv:1903.05958 [hep-ph]}}.
%%CITATION = ARXIV:1903.05958;%%.

\bibitem{Kawamura:2019rth}
J.~Kawamura, S.~Raby, and A.~Trautner, ``{Complete vectorlike fourth family and
  new $U(1)^{\prime}$ for muon anomalies},''
  \href{http://dx.doi.org/10.1103/PhysRevD.100.055030}{{\em Phys. Rev.}
  {\bfseries D100} no.~5, (2019) 055030},
\href{http://arxiv.org/abs/1906.11297}{{\ttfamily arXiv:1906.11297 [hep-ph]}}.
%%CITATION = ARXIV:1906.11297;%%.

\bibitem{Bauer:2019gfk}
M.~Bauer, M.~Neubert, S.~Renner, M.~Schnubel, and A.~Thamm, ``{Axionlike
  Particles, Lepton-Flavor Violation, and a New Explanation of $a_\mu$ and
  $a_e$},'' \href{http://dx.doi.org/10.1103/PhysRevLett.124.211803}{{\em Phys.
  Rev. Lett.} {\bfseries 124} no.~21, (2020) 211803},
\href{http://arxiv.org/abs/1908.00008}{{\ttfamily arXiv:1908.00008 [hep-ph]}}.
%%CITATION = ARXIV:1908.00008;%%.

\bibitem{Han:2018znu}
X.-F. Han, T.~Li, L.~Wang, and Y.~Zhang, ``{Simple interpretations of lepton
  anomalies in the lepton-specific inert two-Higgs-doublet model},''
  \href{http://dx.doi.org/10.1103/PhysRevD.99.095034}{{\em Phys. Rev.}
  {\bfseries D99} no.~9, (2019) 095034},
\href{http://arxiv.org/abs/1812.02449}{{\ttfamily arXiv:1812.02449 [hep-ph]}}.
%%CITATION = ARXIV:1812.02449;%%.

\bibitem{Dutta:2018fge}
B.~Dutta and Y.~Mimura, ``{Electron $g-2$ with flavor violation in MSSM},''
  \href{http://dx.doi.org/10.1016/j.physletb.2018.12.070}{{\em Phys. Lett.}
  {\bfseries B790} (2019) 563--567},
\href{http://arxiv.org/abs/1811.10209}{{\ttfamily arXiv:1811.10209 [hep-ph]}}.
%%CITATION = ARXIV:1811.10209;%%.

\bibitem{Badziak:2019gaf}
M.~Badziak and K.~Sakurai, ``{Explanation of electron and muon $g-2$ anomalies
  in the MSSM},'' \href{http://dx.doi.org/10.1007/JHEP10(2019)024}{{\em JHEP}
  {\bfseries 10} (2019) 024},
\href{http://arxiv.org/abs/1908.03607}{{\ttfamily arXiv:1908.03607 [hep-ph]}}.
%%CITATION = ARXIV:1908.03607;%%.

\bibitem{Hiller:2019mou}
G.~Hiller, C.~Hormigos-Feliu, D.~F. Litim, and T.~Steudtner, ``{Anomalous
  magnetic moments from asymptotic safety},''
  \href{http://dx.doi.org/10.1103/PhysRevD.102.071901}{{\em Phys. Rev.}
  {\bfseries D102} no.~7, (2020) 071901},
\href{http://arxiv.org/abs/1910.14062}{{\ttfamily arXiv:1910.14062 [hep-ph]}}.
%%CITATION = ARXIV:1910.14062;%%.

\bibitem{CarcamoHernandez:2019ydc}
A.~E. Cárcamo~Hernández, S.~F. King, H.~Lee, and S.~J. Rowley, ``{Is it
  possible to explain the muon and electron $g-2$ in a $Z^{\prime}$ model?},''
  \href{http://dx.doi.org/10.1103/PhysRevD.101.115016}{{\em Phys. Rev.}
  {\bfseries D101} no.~11, (2020) 115016},
\href{http://arxiv.org/abs/1910.10734}{{\ttfamily arXiv:1910.10734 [hep-ph]}}.
%%CITATION = ARXIV:1910.10734;%%.

\bibitem{Appelquist:2004mn}
T.~Appelquist, M.~Piai, and R.~Shrock, ``{Lepton dipole moments in extended
  technicolor models},''
  \href{http://dx.doi.org/10.1016/j.physletb.2004.04.062}{{\em Phys. Lett.}
  {\bfseries B593} (2004) 175--180},
\href{http://arxiv.org/abs/hep-ph/0401114}{{\ttfamily arXiv:hep-ph/0401114
  [hep-ph]}}.
%%CITATION = HEP-PH/0401114;%%.

\bibitem{Gerard:1982mm}
J.~M. Gerard, ``{FERMION MASS SPECTRUM IN SU(2)-L x U(1)},''
\href{http://dx.doi.org/10.1007/BF01572477}{{\em Z. Phys.} {\bfseries C18}
  (1983) 145}.
%%CITATION = ZEPYA,C18,145;%%.

\bibitem{Kubo:2003iw}
J.~Kubo, A.~Mondragon, M.~Mondragon, and E.~Rodriguez-Jauregui, ``{The Flavor
  symmetry},'' \href{http://dx.doi.org/10.1143/PTP.109.795}{{\em Prog. Theor.
  Phys.} {\bfseries 109} (2003) 795--807},
  \href{http://arxiv.org/abs/hep-ph/0302196}{{\ttfamily arXiv:hep-ph/0302196
  [hep-ph]}}.
[Erratum: Prog. Theor. Phys.114,287(2005)].
%%CITATION = HEP-PH/0302196;%%.

\bibitem{Kubo:2003pd}
J.~Kubo, ``{Majorana phase in minimal S(3) invariant extension of the standard
  model},'' \href{http://dx.doi.org/10.1016/j.physletb.2005.06.013,
  10.1016/j.physletb.2003.10.048}{{\em Phys. Lett.} {\bfseries B578} (2004)
  156--164}, \href{http://arxiv.org/abs/hep-ph/0309167}{{\ttfamily
  arXiv:hep-ph/0309167 [hep-ph]}}.
[Erratum: Phys. Lett.B619,387(2005)].
%%CITATION = HEP-PH/0309167;%%.

\bibitem{Kobayashi:2003fh}
T.~Kobayashi, J.~Kubo, and H.~Terao, ``{Exact S(3) symmetry solving the
  supersymmetric flavor problem},''
  \href{http://dx.doi.org/10.1016/j.physletb.2003.03.002}{{\em Phys. Lett.}
  {\bfseries B568} (2003) 83--91},
\href{http://arxiv.org/abs/hep-ph/0303084}{{\ttfamily arXiv:hep-ph/0303084
  [hep-ph]}}.
%%CITATION = HEP-PH/0303084;%%.

\bibitem{Chen:2004rr}
S.-L. Chen, M.~Frigerio, and E.~Ma, ``{Large neutrino mixing and normal mass
  hierarchy: A Discrete understanding},''
  \href{http://dx.doi.org/10.1103/PhysRevD.70.079905,
  10.1103/PhysRevD.70.073008}{{\em Phys. Rev.} {\bfseries D70} (2004) 073008},
  \href{http://arxiv.org/abs/hep-ph/0404084}{{\ttfamily arXiv:hep-ph/0404084
  [hep-ph]}}.
[Erratum: Phys. Rev.D70,079905(2004)].
%%CITATION = HEP-PH/0404084;%%.

\bibitem{Mondragon:2007af}
A.~Mondragon, M.~Mondragon, and E.~Peinado, ``{Lepton masses, mixings and FCNC
  in a minimal S(3)-invariant extension of the Standard Model},''
  \href{http://dx.doi.org/10.1103/PhysRevD.76.076003}{{\em Phys. Rev.}
  {\bfseries D76} (2007) 076003},
\href{http://arxiv.org/abs/0706.0354}{{\ttfamily arXiv:0706.0354 [hep-ph]}}.
%%CITATION = ARXIV:0706.0354;%%.

\bibitem{Mondragon:2008gm}
A.~Mondragon, M.~Mondragon, and E.~Peinado, ``{Lepton flavour violating
  processes in an S(3)-symmetric model},'' {\em Rev. Mex. Fis.} {\bfseries 54}
  no.~3, (2008) 81--91, \href{http://arxiv.org/abs/0805.3507}{{\ttfamily
  arXiv:0805.3507 [hep-ph]}}.
[Rev. Mex. Fis. Suppl.54,0181(2008)].
%%CITATION = ARXIV:0805.3507;%%.

\bibitem{Bhattacharyya:2010hp}
G.~Bhattacharyya, P.~Leser, and H.~Pas, ``{Exotic Higgs boson decay modes as a
  harbinger of $S_3$ flavor symmetry},''
  \href{http://dx.doi.org/10.1103/PhysRevD.83.011701}{{\em Phys. Rev.}
  {\bfseries D83} (2011) 011701},
\href{http://arxiv.org/abs/1006.5597}{{\ttfamily arXiv:1006.5597 [hep-ph]}}.
%%CITATION = ARXIV:1006.5597;%%.

\bibitem{Dias:2012bh}
A.~G. Dias, A.~C.~B. Machado, and C.~C. Nishi, ``{An $S_3$ Model for Lepton
  Mass Matrices with Nearly Minimal Texture},''
  \href{http://dx.doi.org/10.1103/PhysRevD.86.093005}{{\em Phys. Rev.}
  {\bfseries D86} (2012) 093005},
\href{http://arxiv.org/abs/1206.6362}{{\ttfamily arXiv:1206.6362 [hep-ph]}}.
%%CITATION = ARXIV:1206.6362;%%.

\bibitem{Meloni:2012ci}
D.~Meloni, ``{$S_3$ as a flavour symmetry for quarks and leptons after the Daya
  Bay result on $\theta_{13}$},''
  \href{http://dx.doi.org/10.1007/JHEP05(2012)124}{{\em JHEP} {\bfseries 05}
  (2012) 124},
\href{http://arxiv.org/abs/1203.3126}{{\ttfamily arXiv:1203.3126 [hep-ph]}}.
%%CITATION = ARXIV:1203.3126;%%.

\bibitem{Canales:2012dr}
F.~Gonzalez~Canales, A.~Mondragon, and M.~Mondragon, ``{The $S_3$ Flavour
  Symmetry: Neutrino Masses and Mixings},''
  \href{http://dx.doi.org/10.1002/prop.201200121}{{\em Fortsch. Phys.}
  {\bfseries 61} (2013) 546--570},
\href{http://arxiv.org/abs/1205.4755}{{\ttfamily arXiv:1205.4755 [hep-ph]}}.
%%CITATION = ARXIV:1205.4755;%%.

\bibitem{Canales:2013cga}
F.~González~Canales, A.~Mondragón, M.~Mondragón, U.~J. Saldaña~Salazar, and
  L.~Velasco-Sevilla, ``{Quark sector of S3 models: classification and
  comparison with experimental data},''
  \href{http://dx.doi.org/10.1103/PhysRevD.88.096004}{{\em Phys. Rev.}
  {\bfseries D88} (2013) 096004},
\href{http://arxiv.org/abs/1304.6644}{{\ttfamily arXiv:1304.6644 [hep-ph]}}.
%%CITATION = ARXIV:1304.6644;%%.

\bibitem{Ma:2013zca}
E.~Ma and B.~Melic, ``{Updated $S_{3}$ model of quarks},''
  \href{http://dx.doi.org/10.1016/j.physletb.2013.07.015}{{\em Phys. Lett.}
  {\bfseries B725} (2013) 402--406},
\href{http://arxiv.org/abs/1303.6928}{{\ttfamily arXiv:1303.6928 [hep-ph]}}.
%%CITATION = ARXIV:1303.6928;%%.

\bibitem{Kajiyama:2013sza}
Y.~Kajiyama, H.~Okada, and K.~Yagyu, ``{Electron/Muon Specific Two Higgs
  Doublet Model},''
  \href{http://dx.doi.org/10.1016/j.nuclphysb.2014.08.009}{{\em Nucl. Phys.}
  {\bfseries B887} (2014) 358--370},
\href{http://arxiv.org/abs/1309.6234}{{\ttfamily arXiv:1309.6234 [hep-ph]}}.
%%CITATION = ARXIV:1309.6234;%%.

\bibitem{Ma:2014qra}
E.~Ma and R.~Srivastava, ``{Dirac or inverse seesaw neutrino masses with $B-L$
  gauge symmetry and $S_3$ flavor symmetry},''
  \href{http://dx.doi.org/10.1016/j.physletb.2014.12.049}{{\em Phys. Lett.}
  {\bfseries B741} (2015) 217--222},
\href{http://arxiv.org/abs/1411.5042}{{\ttfamily arXiv:1411.5042 [hep-ph]}}.
%%CITATION = ARXIV:1411.5042;%%.

\bibitem{Gupta:2014nba}
S.~Gupta, C.~S. Kim, and P.~Sharma, ``{Radiative and seesaw threshold
  corrections to the $S_3$ symmetric neutrino mass matrix},''
  \href{http://dx.doi.org/10.1016/j.physletb.2014.12.005}{{\em Phys. Lett.}
  {\bfseries B740} (2015) 353--358},
\href{http://arxiv.org/abs/1408.0172}{{\ttfamily arXiv:1408.0172 [hep-ph]}}.
%%CITATION = ARXIV:1408.0172;%%.

\bibitem{Hernandez:2015dga}
A.~E. Cárcamo~Hernández, I.~de~Medeiros~Varzielas, and E.~Schumacher,
  ``{Fermion and scalar phenomenology of a two-Higgs-doublet model with
  $S_3$},'' \href{http://dx.doi.org/10.1103/PhysRevD.93.016003}{{\em Phys.
  Rev.} {\bfseries D93} no.~1, (2016) 016003},
\href{http://arxiv.org/abs/1509.02083}{{\ttfamily arXiv:1509.02083 [hep-ph]}}.
%%CITATION = ARXIV:1509.02083;%%.

\bibitem{Hernandez:2015zeh}
A.~E. Cárcamo~Hernández, I.~de~Medeiros~Varzielas, and N.~A. Neill, ``{Novel
  Randall-Sundrum model with $S_{3}$ flavor symmetry},''
  \href{http://dx.doi.org/10.1103/PhysRevD.94.033011}{{\em Phys. Rev.}
  {\bfseries D94} no.~3, (2016) 033011},
\href{http://arxiv.org/abs/1511.07420}{{\ttfamily arXiv:1511.07420 [hep-ph]}}.
%%CITATION = ARXIV:1511.07420;%%.

\bibitem{Hernandez:2015hrt}
A.~E. Cárcamo~Hernández, ``{A novel and economical explanation for SM fermion
  masses and mixings},''
  \href{http://dx.doi.org/10.1140/epjc/s10052-016-4351-y}{{\em Eur. Phys. J.}
  {\bfseries C76} no.~9, (2016) 503},
\href{http://arxiv.org/abs/1512.09092}{{\ttfamily arXiv:1512.09092 [hep-ph]}}.
%%CITATION = ARXIV:1512.09092;%%.

\bibitem{Hernandez:2016rbi}
A.~E. Cárcamo~Hernández, I.~de~Medeiros~Varzielas, and E.~Schumacher, ``{The
  $750\,\text{GeV}$ diphoton resonance in the light of a 2HDM with $S_3$
  flavour symmetry},''
\href{http://arxiv.org/abs/1601.00661}{{\ttfamily arXiv:1601.00661 [hep-ph]}}.
%%CITATION = ARXIV:1601.00661;%%.

\bibitem{CarcamoHernandez:2016pdu}
A.~E. Cárcamo~Hernández, S.~Kovalenko, and I.~Schmidt, ``{Radiatively
  generated hierarchy of lepton and quark masses},''
  \href{http://dx.doi.org/10.1007/JHEP02(2017)125}{{\em JHEP} {\bfseries 02}
  (2017) 125},
\href{http://arxiv.org/abs/1611.09797}{{\ttfamily arXiv:1611.09797 [hep-ph]}}.
%%CITATION = ARXIV:1611.09797;%%.

\bibitem{Arbelaez:2016mhg}
C.~Arbeláez, A.~E. Cárcamo~Hernández, S.~Kovalenko, and I.~Schmidt,
  ``{Radiative Seesaw-type Mechanism of Fermion Masses and Non-trivial Quark
  Mixing},'' \href{http://dx.doi.org/10.1140/epjc/s10052-017-4948-9}{{\em Eur.
  Phys. J.} {\bfseries C77} no.~6, (2017) 422},
\href{http://arxiv.org/abs/1602.03607}{{\ttfamily arXiv:1602.03607 [hep-ph]}}.
%%CITATION = ARXIV:1602.03607;%%.

\bibitem{Gomez-Izquierdo:2017rxi}
J.~C. Gómez-Izquierdo, ``{Non-minimal flavored ${S}_{3}\otimes {Z}_{2}$
  left–right symmetric model},''
  \href{http://dx.doi.org/10.1140/epjc/s10052-017-5094-0}{{\em Eur. Phys. J.}
  {\bfseries C77} no.~8, (2017) 551},
\href{http://arxiv.org/abs/1701.01747}{{\ttfamily arXiv:1701.01747 [hep-ph]}}.
%%CITATION = ARXIV:1701.01747;%%.

\bibitem{Cruz:2017add}
A.~A. Cruz and M.~Mondragón, ``{Neutrino masses, mixing, and leptogenesis in
  an S3 model},''
\href{http://arxiv.org/abs/1701.07929}{{\ttfamily arXiv:1701.07929 [hep-ph]}}.
%%CITATION = ARXIV:1701.07929;%%.

\bibitem{Ma:2017trv}
E.~Ma, ``{Cobimaximal neutrino mixing from $S_3 \times Z_2$},''
  \href{http://dx.doi.org/10.1016/j.physletb.2017.12.049}{{\em Phys. Lett.}
  {\bfseries B777} (2018) 332--334},
\href{http://arxiv.org/abs/1707.03352}{{\ttfamily arXiv:1707.03352 [hep-ph]}}.
%%CITATION = ARXIV:1707.03352;%%.

\bibitem{Espinoza:2018itz}
C.~Espinoza, E.~A. Garcés, M.~Mondragón, and H.~Reyes-González, ``{The $S3$
  Symmetric Model with a Dark Scalar},''
  \href{http://dx.doi.org/10.1016/j.physletb.2018.11.028}{{\em Phys. Lett.}
  {\bfseries B788} (2019) 185--191},
\href{http://arxiv.org/abs/1804.01879}{{\ttfamily arXiv:1804.01879 [hep-ph]}}.
%%CITATION = ARXIV:1804.01879;%%.

\bibitem{Garces:2018nar}
E.~A. Garcés, J.~C. Gómez-Izquierdo, and F.~Gonzalez-Canales, ``{Flavored
  non-minimal left–right symmetric model fermion masses and mixings},''
  \href{http://dx.doi.org/10.1140/epjc/s10052-018-6271-5}{{\em Eur. Phys. J.}
  {\bfseries C78} no.~10, (2018) 812},
\href{http://arxiv.org/abs/1807.02727}{{\ttfamily arXiv:1807.02727 [hep-ph]}}.
%%CITATION = ARXIV:1807.02727;%%.

\bibitem{CarcamoHernandez:2018vdj}
A.~E. Cárcamo~Hernández, J.~Vignatti, and A.~Zerwekh, ``{Generating lepton
  masses and mixings with a heavy vector doublet},''
  \href{http://dx.doi.org/10.1088/1361-6471/ab4499}{{\em J. Phys.} {\bfseries
  G46} no.~11, (2019) 115007},
\href{http://arxiv.org/abs/1807.05321}{{\ttfamily arXiv:1807.05321 [hep-ph]}}.
%%CITATION = ARXIV:1807.05321;%%.

\bibitem{Gomez-Izquierdo:2018jrx}
J.~C. Gómez-Izquierdo and M.~Mondragón, ``{B–L Model with $\mathbf{S}_{3}$
  symmetry: Nearest Neighbor Interaction Textures and Broken
  $\mu\leftrightarrow\tau$ Symmetry},''
  \href{http://dx.doi.org/10.1140/epjc/s10052-019-6785-5}{{\em Eur. Phys. J.}
  {\bfseries C79} no.~3, (2019) 285},
\href{http://arxiv.org/abs/1804.08746}{{\ttfamily arXiv:1804.08746 [hep-ph]}}.
%%CITATION = ARXIV:1804.08746;%%.

\bibitem{Pramanick:2019oxb}
S.~Pramanick, ``{Scotogenic S3 symmetric generation of realistic neutrino
  mixing},'' \href{http://dx.doi.org/10.1103/PhysRevD.100.035009}{{\em Phys.
  Rev.} {\bfseries D100} no.~3, (2019) 035009},
\href{http://arxiv.org/abs/1904.07558}{{\ttfamily arXiv:1904.07558 [hep-ph]}}.
%%CITATION = ARXIV:1904.07558;%%.

\bibitem{Salazar:2015gxa}
C.~Salazar, R.~H. Benavides, W.~A. Ponce, and E.~Rojas, ``{LHC Constraints on
  3-3-1 Models},'' \href{http://dx.doi.org/10.1007/JHEP07(2015)096}{{\em JHEP}
  {\bfseries 07} (2015) 096},
\href{http://arxiv.org/abs/1503.03519}{{\ttfamily arXiv:1503.03519 [hep-ph]}}.
%%CITATION = ARXIV:1503.03519;%%.

\bibitem{Huyen:2012uk}
V.~T.~N. Huyen, H.~N. Long, T.~T. Lam, and V.~Q. Phong, ``{Neutral Current in
  Reduced Minimal 3-3-1 Model},''
  \href{http://dx.doi.org/10.15625/0868-3166/24/2/3774}{{\em Commun.in Phys.}
  {\bfseries 24} no.~2, (2014) 97},
\href{http://arxiv.org/abs/1210.5833}{{\ttfamily arXiv:1210.5833 [hep-ph]}}.
%%CITATION = ARXIV:1210.5833;%%.

\bibitem{Martinez:2008jj}
R.~Martinez and F.~Ochoa, ``{Mass-matrix ansatz and constraints on B0(s) -
  anti-B0(s) mixing in 331 models},''
  \href{http://dx.doi.org/10.1103/PhysRevD.77.065012}{{\em Phys. Rev.}
  {\bfseries D77} (2008) 065012},
\href{http://arxiv.org/abs/0802.0309}{{\ttfamily arXiv:0802.0309 [hep-ph]}}.
%%CITATION = ARXIV:0802.0309;%%.

\bibitem{Buras:2013dea}
A.~J. Buras, F.~De~Fazio, and J.~Girrbach, ``{331 models facing new $b \to
  s\mu^+ \mu^-$ data},'' \href{http://dx.doi.org/10.1007/JHEP02(2014)112}{{\em
  JHEP} {\bfseries 02} (2014) 112},
\href{http://arxiv.org/abs/1311.6729}{{\ttfamily arXiv:1311.6729 [hep-ph]}}.
%%CITATION = ARXIV:1311.6729;%%.

\bibitem{Buras:2014yna}
A.~J. Buras, F.~De~Fazio, and J.~Girrbach-Noe, ``{$Z$-$Z'$ mixing and
  $Z$-mediated FCNCs in $SU(3)_{C} \times SU(3)_{L} \times U(1)_{X}$ models},''
  \href{http://dx.doi.org/10.1007/JHEP08(2014)039}{{\em JHEP} {\bfseries 08}
  (2014) 039},
\href{http://arxiv.org/abs/1405.3850}{{\ttfamily arXiv:1405.3850 [hep-ph]}}.
%%CITATION = ARXIV:1405.3850;%%.

\bibitem{Buras:2012dp}
A.~J. Buras, F.~De~Fazio, J.~Girrbach, and M.~V. Carlucci, ``{The Anatomy of
  Quark Flavour Observables in 331 Models in the Flavour Precision Era},''
  \href{http://dx.doi.org/10.1007/JHEP02(2013)023}{{\em JHEP} {\bfseries 02}
  (2013) 023},
\href{http://arxiv.org/abs/1211.1237}{{\ttfamily arXiv:1211.1237 [hep-ph]}}.
%%CITATION = ARXIV:1211.1237;%%.

\bibitem{Kubo:2004ps}
J.~Kubo, H.~Okada, and F.~Sakamaki, ``{Higgs potential in minimal S(3)
  invariant extension of the standard model},''
  \href{http://dx.doi.org/10.1103/PhysRevD.70.036007}{{\em Phys. Rev.}
  {\bfseries D70} (2004) 036007},
\href{http://arxiv.org/abs/hep-ph/0402089}{{\ttfamily arXiv:hep-ph/0402089
  [hep-ph]}}.
%%CITATION = HEP-PH/0402089;%%.

\bibitem{Hagiwara:2011af}
K.~Hagiwara, R.~Liao, A.~D. Martin, D.~Nomura, and T.~Teubner, ``{$(g-2)_\mu$
  and $\alpha(M^2_Z)$ re-evaluated using new precise data},''
  \href{http://dx.doi.org/10.1088/0954-3899/38/8/085003}{{\em J. Phys.}
  {\bfseries G38} (2011) 085003},
\href{http://arxiv.org/abs/1105.3149}{{\ttfamily arXiv:1105.3149 [hep-ph]}}.
%%CITATION = ARXIV:1105.3149;%%.

\bibitem{Davier:2017zfy}
M.~Davier, A.~Hoecker, B.~Malaescu, and Z.~Zhang, ``{Reevaluation of the
  hadronic vacuum polarisation contributions to the Standard Model predictions
  of the muon $g-2$ and ${\alpha (m_Z^2)}$ using newest hadronic cross-section
  data},'' \href{http://dx.doi.org/10.1140/epjc/s10052-017-5161-6}{{\em Eur.
  Phys. J.} {\bfseries C77} no.~12, (2017) 827},
\href{http://arxiv.org/abs/1706.09436}{{\ttfamily arXiv:1706.09436 [hep-ph]}}.
%%CITATION = ARXIV:1706.09436;%%.

\bibitem{Nomura:2018lsx}
T.~Nomura and H.~Okada, ``{One-loop neutrino mass model without any additional
  symmetries},'' \href{http://dx.doi.org/10.1016/j.dark.2019.100359}{{\em Phys.
  Dark Univ.} {\bfseries 26} (2019) 100359},
\href{http://arxiv.org/abs/1808.05476}{{\ttfamily arXiv:1808.05476 [hep-ph]}}.
%%CITATION = ARXIV:1808.05476;%%.

\bibitem{Nomura:2018vfz}
T.~Nomura and H.~Okada, ``{Zee-Babu type model with $U(1)_{L_\mu - L_\tau}$
  gauge symmetry},'' \href{http://dx.doi.org/10.1103/PhysRevD.97.095023}{{\em
  Phys. Rev.} {\bfseries D97} no.~9, (2018) 095023},
\href{http://arxiv.org/abs/1803.04795}{{\ttfamily arXiv:1803.04795 [hep-ph]}}.
%%CITATION = ARXIV:1803.04795;%%.

\bibitem{Blum:2018mom}
{\bfseries RBC, UKQCD} Collaboration, T.~Blum, P.~A. Boyle, V.~Gülpers,
  T.~Izubuchi, L.~Jin, C.~Jung, A.~Jüttner, C.~Lehner, A.~Portelli, and J.~T.
  Tsang, ``{Calculation of the hadronic vacuum polarization contribution to the
  muon anomalous magnetic moment},''
  \href{http://dx.doi.org/10.1103/PhysRevLett.121.022003}{{\em Phys. Rev.
  Lett.} {\bfseries 121} no.~2, (2018) 022003},
\href{http://arxiv.org/abs/1801.07224}{{\ttfamily arXiv:1801.07224 [hep-lat]}}.
%%CITATION = ARXIV:1801.07224;%%.

\bibitem{Keshavarzi:2018mgv}
A.~Keshavarzi, D.~Nomura, and T.~Teubner, ``{Muon $g-2$ and $\alpha(M_Z^2)$: a
  new data-based analysis},''
  \href{http://dx.doi.org/10.1103/PhysRevD.97.114025}{{\em Phys. Rev.}
  {\bfseries D97} no.~11, (2018) 114025},
\href{http://arxiv.org/abs/1802.02995}{{\ttfamily arXiv:1802.02995 [hep-ph]}}.
%%CITATION = ARXIV:1802.02995;%%.

\bibitem{Aoyama:2020ynm}
T.~Aoyama {\em et~al.}, ``{The anomalous magnetic moment of the muon in the
  Standard Model},''
  \href{http://dx.doi.org/10.1016/j.physrep.2020.07.006}{{\em Phys. Rept.}
  {\bfseries 887} (2020) 1--166},
\href{http://arxiv.org/abs/2006.04822}{{\ttfamily arXiv:2006.04822 [hep-ph]}}.
%%CITATION = ARXIV:2006.04822;%%.

\bibitem{Parker:2018vye}
R.~H. Parker, C.~Yu, W.~Zhong, B.~Estey, and H.~Müller, ``{Measurement of the
  fine-structure constant as a test of the Standard Model},''
  \href{http://dx.doi.org/10.1126/science.aap7706}{{\em Science} {\bfseries
  360} (2018) 191},
\href{http://arxiv.org/abs/1812.04130}{{\ttfamily arXiv:1812.04130
  [physics.atom-ph]}}.
%%CITATION = ARXIV:1812.04130;%%.

\bibitem{Grimus:2000vj}
W.~Grimus and L.~Lavoura, ``{The Seesaw mechanism at arbitrary order:
  Disentangling the small scale from the large scale},''
  \href{http://dx.doi.org/10.1088/1126-6708/2000/11/042}{{\em JHEP} {\bfseries
  11} (2000) 042},
\href{http://arxiv.org/abs/hep-ph/0008179}{{\ttfamily arXiv:hep-ph/0008179
  [hep-ph]}}.
%%CITATION = HEP-PH/0008179;%%.

\bibitem{Diaz:2002uk}
R.~A. Diaz, R.~Martinez, and J.~A. Rodriguez, ``{Phenomenology of lepton flavor
  violation in 2HDM(3) from (g-2)(mu) and leptonic decays},''
  \href{http://dx.doi.org/10.1103/PhysRevD.67.075011}{{\em Phys. Rev.}
  {\bfseries D67} (2003) 075011},
\href{http://arxiv.org/abs/hep-ph/0208117}{{\ttfamily arXiv:hep-ph/0208117
  [hep-ph]}}.
%%CITATION = HEP-PH/0208117;%%.

\bibitem{Lindner:2016bgg}
M.~Lindner, M.~Platscher, and F.~S. Queiroz, ``{A Call for New Physics : The
  Muon Anomalous Magnetic Moment and Lepton Flavor Violation},''
  \href{http://dx.doi.org/10.1016/j.physrep.2017.12.001}{{\em Phys. Rept.}
  {\bfseries 731} (2018) 1--82},
\href{http://arxiv.org/abs/1610.06587}{{\ttfamily arXiv:1610.06587 [hep-ph]}}.
%%CITATION = ARXIV:1610.06587;%%.

\bibitem{Kowalska:2017iqv}
K.~Kowalska and E.~M. Sessolo, ``{Expectations for the muon g-2 in simplified
  models with dark matter},''
  \href{http://dx.doi.org/10.1007/JHEP09(2017)112}{{\em JHEP} {\bfseries 09}
  (2017) 112},
\href{http://arxiv.org/abs/1707.00753}{{\ttfamily arXiv:1707.00753 [hep-ph]}}.
%%CITATION = ARXIV:1707.00753;%%.

\bibitem{Bora:2012tx}
K.~Bora, ``{Updated values of running quark and lepton masses at GUT scale in
  SM, 2HDM and MSSM},'' {\em Horizon} {\bfseries 2} (2013) 112,
\href{http://arxiv.org/abs/1206.5909}{{\ttfamily arXiv:1206.5909 [hep-ph]}}.
%%CITATION = ARXIV:1206.5909;%%.

\bibitem{Xing:2007fb}
Z.-z. Xing, H.~Zhang, and S.~Zhou, ``{Updated Values of Running Quark and
  Lepton Masses},'' \href{http://dx.doi.org/10.1103/PhysRevD.77.113016}{{\em
  Phys. Rev.} {\bfseries D77} (2008) 113016},
\href{http://arxiv.org/abs/0712.1419}{{\ttfamily arXiv:0712.1419 [hep-ph]}}.
%%CITATION = ARXIV:0712.1419;%%.

\bibitem{Tanabashi:2018oca}
{\bfseries Particle Data Group} Collaboration, M.~Tanabashi {\em et~al.},
  ``{Review of Particle Physics},''
\href{http://dx.doi.org/10.1103/PhysRevD.98.030001}{{\em Phys. Rev.} {\bfseries
  D98} no.~3, (2018) 030001}.
%%CITATION = PHRVA,D98,030001;%%.

\bibitem{Ma:2006km}
E.~Ma, ``{Verifiable radiative seesaw mechanism of neutrino mass and dark
  matter},'' \href{http://dx.doi.org/10.1103/PhysRevD.73.077301}{{\em Phys.
  Rev.} {\bfseries D73} (2006) 077301},
\href{http://arxiv.org/abs/hep-ph/0601225}{{\ttfamily arXiv:hep-ph/0601225
  [hep-ph]}}.
%%CITATION = HEP-PH/0601225;%%.

\bibitem{Ishimori:2010au}
H.~Ishimori, T.~Kobayashi, H.~Ohki, Y.~Shimizu, H.~Okada, and M.~Tanimoto,
  ``{Non-Abelian Discrete Symmetries in Particle Physics},''
  \href{http://dx.doi.org/10.1143/PTPS.183.1}{{\em Prog. Theor. Phys. Suppl.}
  {\bfseries 183} (2010) 1--163},
\href{http://arxiv.org/abs/1003.3552}{{\ttfamily arXiv:1003.3552 [hep-th]}}.
%%CITATION = ARXIV:1003.3552;%%.

\end{thebibliography}\endgroup

\end{document}